\documentclass[12pt,preprint]{aastex}

\usepackage{natbib}
\usepackage{graphicx}
\usepackage{url}

\newcommand{\Hz}{\textrm{Hz}}
\newcommand{\kHz}{\textrm{kHz}}
\newcommand{\MHz}{\textrm{MHz}}
\newcommand{\GHz}{\textrm{GHz}}
\newcommand{\gm}{\textrm{gm}}
\newcommand{\cm}{\textrm{cm}}
\newcommand{\m}{\textrm{m}}
\newcommand{\km}{\textrm{km}}
\newcommand{\kpc}{\textrm{kpc}} 
\newcommand{\us}{\ensuremath\mathrm{\mu s}}
\newcommand{\ms}{\textrm{ms}}
\newcommand{\s}{\textrm{s}} 
\newcommand{\hr}{\textrm{hr}}
\newcommand{\pyr}{\ensuremath{\mathrm{yr^{-1}}}}
\newcommand{\pmsq}{\ensuremath{\mathrm{m^{-2}}}}
\newcommand{\ps}{\ensuremath{\mathrm{s^{-1}}}}
\newcommand{\K}{\textrm{K}}
\newcommand{\gauss}{\textrm{G}}
\newcommand{\erg}{\textrm{erg}}
\newcommand{\Jy}{\textrm{Jy}}
\newcommand{\mJy}{\textrm{mJy}}
\newcommand{\Msun}{\ensuremath{\mathrm{M_{\Sun}}}}
\newcommand{\dm}{\textrm{DM}}
\newcommand{\dmu}{\ensuremath{\mathrm{pc\; cm^{-3}}}}
\newcommand{\sn}{\textrm{S/N}} 

\newcommand{\rmsub}[1]{\ensuremath{_{\mathrm{#1}}}}

\defcitealias{dss}{Paper I}

\title{The Green Bank Telescope $\mathbf{350\; \MHz}$ Drift-scan
  Survey II: Data Analysis and the Timing of 10 New Pulsars, Including
  a Relativistic Binary}

\shorttitle{The GBT Drift-scan Survey II}

\author{Ryan S.\ Lynch\altaffilmark{1,2}, Jason
  Boyles\altaffilmark{3,4}, Scott M.\ Ransom\altaffilmark{5}, Ingrid
  H.\ Stairs\altaffilmark{6}, Duncan R.\ Lorimer\altaffilmark{3,7},
  Maura A.\ McLaughlin\altaffilmark{3}, Jason W.\ T.\
  Hessels\altaffilmark{8,9}, Victoria M.\ Kaspi\altaffilmark{1},
  Vladislav I.\ Kondratiev\altaffilmark{8,10}, Anne M.\
  Archibald\altaffilmark{1}, Aaron Berndsen\altaffilmark{6}, Rogerio
  F. Cardoso\altaffilmark{3}, Angus Cherry\altaffilmark{6}, Courtney
  R.\ Epstein\altaffilmark{11}, Chen Karako-Argaman\altaffilmark{1},
  Christie A.\ McPhee\altaffilmark{6}, Tim Pennucci\altaffilmark{2},
  Mallory S.\ E.\ Roberts\altaffilmark{12,13}, Kevin
  Stovall\altaffilmark{14,15}, and Joeri van
  Leeuwen\altaffilmark{8,9}}

\altaffiltext{1}{Department of Physics, McGill University, 3600
  University Street, Montreal, QC H3A 2T8, Canada,
  \texttt{rlynch@physics.mcgill.ca}}
\altaffiltext{2}{Department of Astronomy, University of Virginia,
  P.O.\ Box 400325, Charlottesville, VA 22904, USA}
\altaffiltext{3}{Department of Physics, West Virginia University, 111
  White Hall, Morgantown, WV 26506, USA}
\altaffiltext{4}{Department of Physics and Astronomy, Western Kentucky
  University, Bowling Green, KY 42101, USA}
\altaffiltext{5}{National Radio Astronomy Observatory, 520 Edgemont
  Road, Charlottesville, VA 22903, USA}
\altaffiltext{6}{Department of Physics and Astronomy, University of
  British Columbia, 6224 Agricultural Road, Vancouver, BC V6T 1Z1,
  Canada}
\altaffiltext{7}{Also adjunct at the National Radio Astronomy
  Observatory, Green Bank, WV, 24944, USA}
\altaffiltext{8}{ASTRON, the Netherlands Institute for Radio Astronomy,
  Postbus 2, 7990 AA Dwingeloo, The Netherlands}
\altaffiltext{9}{Astronomical Institute ``Anton Pannekoek'', University
  of Amsterdam, Science Park 904, 1098 XH Amsterdam, The Netherlands}
\altaffiltext{10}{Astro Space Center of the Lebedev Physical Institute,
  Profsoyuznaya str. 84/32, Moscow 117997, Russia}
\altaffiltext{11}{Department of Astronomy, Ohio State University, 140
  W.\ 18$^\mathrm{th}$ Avenue, Columbus, OH, 43210, USA}
\altaffiltext{12}{Eureka Scientific Inc., 2452 Delmer Street, Suite
  100, Oakland, CA 94602, USA}
\altaffiltext{13}{Department of Physics, Ithaca College, Ithaca, NY
  14850, USA}
\altaffiltext{14}{Center for Advanced Radio Astronomy and Department
  of Physics and Astronomy, University of Texas at Brownsville,
  Brownsville, TX 78520, USA}
\altaffiltext{15}{Department of Physics and Astronomy, University of
  Texas at San Antonio, San Antonio, TX 78249, USA}

\shortauthors{Lynch et al.}


\begin{document}

\setcounter{footnote}{14}

\begin{abstract}

  We have completed a $350\; \MHz$ drift scan survey using the Robert
  C.\ Byrd Green Bank Telescope with the goal of finding new radio
  pulsars, especially millisecond pulsars that can be timed to high
  precision.  This survey covered $\sim$10300 square degrees and all
  of the data have now been fully processed.  We have discovered a
  total of 31 new pulsars, seven of which are recycled pulsars.  A
  companion paper by \citet{dss} describes the survey strategy, sky
  coverage, and instrumental set-up, and presents timing solutions for
  the first 13 pulsars.  Here we describe the data analysis pipeline,
  survey sensitivity, and follow-up observations of new pulsars, and
  present timing solutions for 10 other pulsars.  We highlight several
  sources---two interesting nulling pulsars, an isolated millisecond
  pulsar with a measurement of proper motion, and a partially recycled
  pulsar, PSR J0348+0432, which has a white dwarf companion in a
  relativistic orbit.  PSR J0348+0432 will enable unprecedented tests
  of theories of gravity.

\end{abstract}

\maketitle
\section{Introduction \label{sec:intro}}

The vast majority of observed neutron stars in the Galaxy manifest
themselves as radio pulsars.  The extremely high rotational stability
of pulsars, and especially millisecond pulsars (MSPs), make them
unrivaled laboratories for studying a wide range of astrophysical
phenomena.  Most pulsars have been discovered in large-area surveys,
but most of these have focused on southern declinations or narrow
regions around the Galactic plane.  There is a need to find more
pulsars in the northern sky, particularly high-precision MSPs that can
be included in a pulsar timing array to detect gravitational waves
\citep[e.g.][]{jfl+09}.  The $100$-$\m$ Robert C.\ Byrd Green Bank
Telescope (GBT) is one of the best telescopes in the world for finding
and studying pulsars and a visible-sky pulsar survey using the GBT is
underway (the Green Bank North Celestial Cap survey).

During the northern summer of 2007 the azimuth track of the GBT
underwent repair, making normal operations impossible\footnote{A
  history of the track repair can be found at
  \url{http://www.gb.nrao.edu/gbt/track.shtml}}.  Our team took
advantage of the situation by completing the GBT $350\; \MHz$
Drift-scan Survey between May and August.  Because the GBT was unable
to move in azimuth, we observed at a number of fixed elevations and
allowed the sky to drift through the telescope beam at the sidereal
rate.  This survey was one of several low-frequency GBT surveys that
are optimized for finding bright, nearby pulsars, with an emphasis on
MSPs.  These surveys have either been completed \citep{hrk+08}, are
ongoing, or are planned for the future.  We collected over 1491 hours
of data totaling $134\; \mathrm{TB}$.  Approximately $30\;
\mathrm{TB}$ of these data are being analyzed by the Pulsar Search
Collaboratory\footnote{\url{http://www.pulsarsearchcollaboratory.com/}},
an educational initiative that actively involves high school students
and teachers in research under the guidance of a team of astronomers
\citep{rhm+10}.  For the remainder of this paper we discuss only the
$\sim$$100\; \mathrm{TB}$ of the data that we have analyzed ourselves.
All of these data have been fully processed and we have discovered 31
new pulsars, including ten recycled pulsars (seven of which are MSPs
with $P < 10\; \ms$).  We have derived full timing solutions for 25 of
these new pulsars.  The first 13 pulsars are being presented in a
companion paper \citep[hereafter \citetalias{dss}]{dss}, along with a
detailed description of the survey strategy, sky coverage, and
instrumental set-up.  We present timing solutions for an additional 10
pulsars here and describe the survey pipeline and data analysis in
detail.  An earlier-discovered Drift-scan pulsar, PSR J1023$+$0038,
has been discussed elsewhere \citep[see][]{asr+08,akb+10}, while
another MSP, PSR J2256$-$1024, will be presented in a future paper
(Stairs et al. in prep).  In \S\ref{sec:analysis} we explain how the
data were divided into regions on the sky, interference removal, our
de-dispersion scheme, and search strategies.  In \S\ref{sec:sens} we
describe our approximate survey sensitivity and the effects of
scattering.  In \S\ref{sec:cand} we describe how we confirmed
candidate pulsars and our follow-up observations.  In
\S\ref{sec:results} we present timing solutions and discuss some
interesting individual systems.  A summary can be found in
\S\ref{sec:conc}.

\section{Data Analysis \label{sec:analysis}}

The Drift-scan survey covered $\sim$10300 square degrees.  A detailed
description of the survey strategy, sky coverage, and instrumental
set-up can be found in \citetalias{dss}.  Here we focus on the data
reduction and search techniques.  All data were processed using the
\texttt{PRESTO}\footnote{\url{http://www.cv.nrao.edu/~sransom/presto/}}
software suite \citep{ran01}.

\subsection{Pseudo-Pointings and Interference Excision
  \label{sec:point_rfi}}

Data were collected while the azimuth track of the GBT was being
repaired, so the telescope was locked at constant azimuth.  Different
regions of the sky were observed by changing the elevation of the
telescope and allowing the sky to drift through the telescope beam at
the sidereal rate.  The time for a given point on the sky to pass
through the beam is
\begin{eqnarray}
  t \simeq \frac{b}{\mathcal{R}\rmsub{sid} \cos{\delta}},
\end{eqnarray}
where $b = 36'$ is the full-width-at-half-maximum of the GBT beam at
$350\; \MHz$, $\mathcal{R}\rmsub{sid} = 15'\, \min^{-1}$ is the
sidereal rate, and $\delta$ is the declination.  The survey covered
$-8^\circ \lesssim \delta \lesssim +38^\circ$ and $-21^\circ \lesssim
\delta \lesssim +38^\circ$, depending on the azimuth of the telescope
(see \citetalias{dss} for details).  Although the telescope was not
actually tracking the sky, we defined an individual
\emph{pseudo-pointing} to be a continuous block of data $\sim$$140\;
\s$ in duration.  Each pseudo-pointing overlapped with the preceding
one by $70\; \s$, so that all of our data were processed as part of
two different pseudo-pointings.

The raw data were collected using the GBT Pulsar Spigot back-end
\citep{kel+05}.  The Spigot uses autocorrelation chips that each work
on 3-level raw samples and create an adjustable number of lags.  These
are then integrated into either 8- or 16-bit values, depending on the
mode.  The center frequency of the observations was $350\; \MHz$ and
the bandwidth was $50\; \MHz$.  Most of our observations were made
using the 8-bit mode that split the band into 2048 lags which were
fast Fourier transformed to synthesize 2048 frequency channels, each
with a width of $24.4\; \kHz$, and recorded every $81.92\; \us$.  A
relatively small amount of data was taken early in the survey using
the 16-bit, 1024 channel mode with the same sampling time.  Hence,
each $140\; \s$ block of data consisted of roughly 1.7 million
spectra.

Each pseudo-pointing was independently analyzed for radio frequency
interference (RFI) using the \texttt{rfifind} tool from
\texttt{PRESTO}.  Data were broken into blocks roughly two seconds
long while maintaining the full frequency resolution.  The total
power, mean, and variance in both the time and frequency domain were
calculated for each data block and compared to the median quantity for
the entire pseudo-pointing.  A time/frequency block was flagged as RFI
and masked out (i.e. set to zero) in future analysis if the value of
total power, mean power, or variance was greater than ten/four
standard deviations from the median values for the entire
pseudo-pointing.  If greater than 30\%/70\% of time intervals/channels
were flagged then all remaining intervals/channels were masked out as
well, under the assumption that they probably contained RFI just below
our cut-off threshold.  In addition to blindly searching for RFI, the
Fourier spectra were de-reddened and persistent, well known sources of
interference (e.g.  the $60\; \Hz$ signal from AC power sources) were
explicitly removed from the power spectra of each pseudo-pointing.
Despite the fact that there was significant construction on-site due
to the track repair, our data were remarkably free of RFI.  For
example, the median masking fraction was 0.56\% and only 0.2\% of the
data had a masking fraction greater than 30\%.  We are confident that
we reached the noise limit for the vast majority of our survey (see
\S3).

\subsection{De-Dispersion \label{sec:dedisp}}

Free electrons in the interstellar medium give rise to a
frequency-dependent dispersive time delay which, if left uncorrected,
will make it virtually impossible to find new pulsars.  The magnitude
of the delay between two frequencies, $\nu_1$ and $\nu_2$, is
\begin{eqnarray}
  t_{\mathrm{DM}} \simeq 4.15 \times 10^3\; \s \times \dm
  \times \left [ \left (\frac{\nu_1}{\MHz} \right )^{-2}
    - \left (\frac{\nu_2}{\MHz} \right )^{-2} \right ],
\end{eqnarray}
where \dm\ is the dispersion measure in units of \dmu.  After applying
appropriate shifts to each frequency channel, we summed over frequency
to create de-dispersed time series.  Each time series was transformed
to the Solar System barycenter using the DE200 ephemeris (the default
used by \texttt{PRESTO}) and the dispersion delay was removed (i.e.,
as if the signal had infinite frequency).  We note that we used the
DE405 ephemeris for baycentering when deriving pulse times of arrival
(see \S\ref{sec:timing}).

The finite size of a frequency channel will still induce a smearing
given by
\begin{eqnarray}
\label{eqn:dm_smear}
t\rmsub{chan} \simeq 8.3 \times 10^3\; \s 
\left (\frac{\dm \Delta \nu\rmsub{chan}}
  {\nu^3} \right ),
\end{eqnarray}
where $\Delta \nu\rmsub{chan}$ is the channel width and $\nu$ is the
channel center frequency (both in megahertz).  For our primary
observing mode with $24.4\; \kHz$ channels centered between $325$ and
$350\; \MHz$, $t_{\mathrm{chan}} \approx (3.8$ to $5.9)\; \us \times
\dm$.

Since the \dm\ of a pulsar is not known \emph{a priori}, we created
de-dispersed time series for each pseudo-pointing over a range of {\dm
  s}, from $0$ to $\sim$$1000\; \dmu$, which is a factor of three to
four larger than the maximum DMs predicted by the NE2001 model
\citep{cl02} in the low Galactic latitude regions of the survey.  The
step size between subsequent trial DMs ($\Delta \dm$) was chosen such
that over the entire band $t_{\Delta \mathrm{DM}} \simeq
t\rmsub{chan}$. This ensures that the maximum extra smearing caused by
any trial DM deviating from the source DM by $\Delta \dm$ is less than
the intra-channel smearing\footnote{De-dispersion plans were generated
  using the \texttt{DDplan.py} tool in \texttt{PRESTO}.}.  To increase
computational efficiency, the data were down-sampled in time by adding
$2^n$ samples together (where $n$ is an integer) when $t\rmsub{chan}
\geq 2^n \times 81.92\; \us$.

\subsection{Search Algorithms \label{sec:searches}}

\subsubsection{Periodic Sources \label{sec:periodic}}

Each RFI-cleaned, de-dispersed time series was Fourier transformed and
searched for periodic signals.  As mentioned in \S\ref{sec:point_rfi},
known sources of RFI were explicitly removed from the Fourier
spectrum, so there is a very small chance ($\sim$$0.006\%$) that a
pulsar with a spin frequency very close to known RFI could have also
been removed.  Acceleration searches for isolated and binary pulsars
were carried out in the Fourier domain \citep{rem00} for signals with
a maximum drift of $z\rmsub{max} = \pm 50/n\rmsub{harm}$ Fourier bins,
where $n\rmsub{harm}$ is the highest harmonic where the pulsar is
detected.  This corresponds to a physical acceleration of
\begin{eqnarray}
  A\rmsub{max} = \frac{z\rmsub{max} c P}{n\rmsub{harm} t\rmsub{int}^2},
  \label{eq:amax}
\end{eqnarray}
where $c$ is the speed of light, $P$ is the spin period of the pulsar,
and $t\rmsub{int} = 140\; \s$ is the effective integration time
\citep{rem00}.  For a $P = 2\; \ms$ pulsar detected with up to eight
harmonics, $A\rmsub{max} \approx 24\; \m\, \s^{-2}$.  Acceleration
searches used up to eight summed harmonics, but we also carried out
searches for unaccelerated pulsars ($z = 0$) using up to 16 summed
harmonics.  Only powers-of-two numbers of harmonics were summed.

To filter out spurious and low-significance signals, only candidates
that appeared in at least two time series of different DMs passed to
the next stage of consideration.  We also filtered duplicate signals
(keeping only those with the highest \sn) that were within $\pm 1.1$
Fourier bins of each other in different DM time series, as well as
those that were harmonically related to each other.  We folded up to
20 of the remaining candidates from the zero-acceleration searches and
up to ten from the high-acceleration searches if their Fourier power
was at least $6 \sigma$ above the Gaussian-equivalent noise level.  We
used the \texttt{prepfold} routine in \texttt{PRESTO} to fold the full
resolution data at the nominal $P$, period derivative ($\dot{P}$), and
\dm\ as determined by our searches.  Our folding code refined these
values and created diagnostic plots that were then saved for human
inspection.

\subsubsection{Single-pulse Sources \label{sec:transient}}

We searched for bright single pulses using
\texttt{single\_pulse\_search.py} in \texttt{PRESTO}.  Each time
series was smoothed using a piecewise linear fit to the data, where
each piece was 2000 points long.  The smoothed data were then
correlated with boxcar functions of varying widths\footnote{The boxcar
functions had a maximum width of either $150 \times n \times dt$ or
$0.1\, \s$, whichever was greater, where $n$ is the down-sampling
factor.}, which acted as matched filters to individual pulses.  We
recorded all single pulses with a signal-to-noise ratio, $\sn \geq 5$
and created diagnostic plots for all pulses with $\sn \geq 5.5$. These
plots were then saved for human inspection.  In addition, an automated
algorithm was used to flag pseudo-pointings with promising candidates,
which were then inspected in greater detail.  Five rotating radio
transients (RRATs) have been discovered in this survey and a further
26 candidates have been identified and await confirmation.  These
discoveries and the automated algorithm used to help identify them
will be presented in a forthcoming paper (Karako-Argaman et al. in
prep).

\section{Survey Sensitivity \label{sec:sens}}

Following \citet{dtws85} and \citet{lk05}, the sensitivity of a pulsar
survey may be written in terms of the phase-averaged limiting flux
density
\begin{eqnarray}
  S\rmsub{\nu,min} = \frac{\beta}{\epsilon} \frac{(\sn\rmsub{min}) 
                     T\rmsub{sys}}{G \sqrt{n\rmsub{p} t\rmsub{int} 
                     \Delta \nu}} \sqrt{\frac{W}{P-W}},
\end{eqnarray}
where $\epsilon$ is a degradation factor (discussed below), $\beta =
1.16$ is a correction factor that accounts for digitization losses,
$\sn\rmsub{min}$ is the \sn\ threshold, $T\rmsub{sys}$ is the total
system temperature, $G$ is the telescope gain, $n\rmsub{p}$ is the
number of summed polarizations, $t\rmsub{int}$ is the integration
time, $\Delta \nu$ is the bandwidth, and $W$ is the total pulse width
(see Table \ref{table:surv_params} for relevant values).  The
degradation factor $\epsilon$ accounts for drift of the pulsar through
the telescope beam, which is not uniform in sensitivity.  For a
Gaussian primary beam
\begin{eqnarray}
\epsilon \propto \int_0^{t\rmsub{int}}{e^{-r^2(t)/f^2}\; dt},
\end{eqnarray}
where $r(t)$ is the distance from the beam center and $f = b/(2
\sqrt{\ln{2}})$.  From simple geometry, $r^2(t) = y^2 + (b/2 - \dot{x}
t)^2$, where $y$ and $x$ are the distances from the beam center in
right ascension and declination, respectively, and $\dot{x}$ is the
drift rate.  We normalize $\epsilon$ such that a pulsar at the center
of the beam \emph{for an entire integration} has $\epsilon = 1$.  For
reference, a pulsar that crosses the beam center will have $\epsilon =
0.81$.

\begin{figure}
\centering
\includegraphics[width=5.75in]{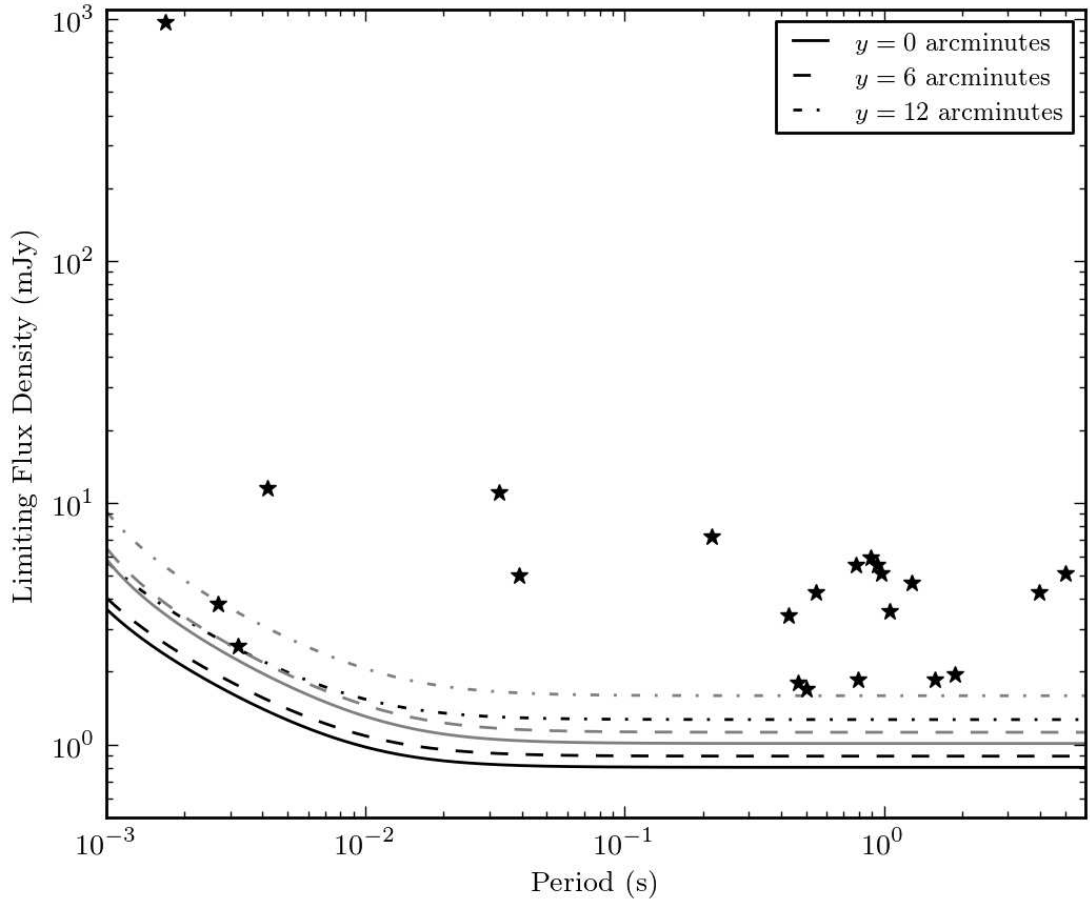}
\caption{Approximate phase-averaged limiting flux density of our
  survey.  Black curves are for $T\rmsub{sys} = 75\; \K$ and $\dm =
  30\; \dmu$, while gray curves are for $T\rmsub{sys} = 100\; \K$ and
  $\dm = 75\; \dmu$.  The smallest offset from the center of the
  telescope beam is $y$.  \label{fig:sens}}
\end{figure}

The system temperature is a sum of several factors, including the
receiver temperature ($T\rmsub{rec}$) and the sky temperature
($T\rmsub{sky}$).  The $350\; \MHz$ receiver\footnote{We have folded
  the receiver spillover and cosmic microwave background into this
  number.  Characteristics of the GBT receivers are available on the
  NRAO website (\url{http://www.gb.nrao.edu/astronomers.shtml}).} of
the GBT has a nominal $T\rmsub{rec} = 23\; \K$.  The Galactic
synchrotron emission contributes heavily to $T\rmsub{sky}$, but this
depends on sky position.  Most of our survey was at high Galactic
latitudes, where the synchrotron emission adds $\sim$30--$50\; \K$ at
$350\; \MHz$ \citep{hss+82}.  Our line of sight through the Galaxy
also affects our sensitivity by increasing scattering and dispersion,
both of which contribute to the observed pulse width.  The typical
maximum predicted \dm\ at the high Galactic latitudes we cover is
$\sim$$60\; \dmu$.  According to the NE2001 model, this corresponds to
a scattering time $\sim$$0.08\; \ms$ at $350\; \MHz$, though observed
scattering times may differ from predictions by an order of magnitude
or more.  Obviously, DM effects become much worse at low Galactic
latitudes.  Figure \ref{fig:sens} shows approximate sensitivity curves
for various combinations of $y$ (the minimum offset from the beam
center), $T\rmsub{sys}$, and \dm.  These calculations do not take the
effects of RFI into account, but as we describe in \S2.1, the survey
did not suffer greatly from RFI contamination.

\section{Candidate Confirmation and Follow-up \label{sec:cand}}

Periodic and single-pulse candidates from each pseudo-pointing were
judged by eye.  Folded candidates were usually judged on three main
criteria: \textit{i}) distinct peaking of the signal's significance at
{\dm s} greater than $0\; \dmu$; \textit{ii}) broad-band emission
(allowing for the possibility of regions of enhanced/diminished flux
due to interstellar scintillation); and \textit{iii}) fairly
persistent emission in time (allowing for eclipses and nulls and
accounting for the roll-off in sensitivity near the edges of the
telescope beam).  In the case of single-pulse candidates, we looked
for pulses that peaked at a non-zero \dm\ and that decreased in
significance away from this peak.  Multiple single pulses at the same
\dm\ were also an obvious indicator of a good candidate.

Promising candidates were confirmed in follow-up observations with the
GBT, after which we began regular timing observations.  To improve the
quality of initial timing solutions, new pulsars had their sky
positions refined by observing at a grid of locations with smaller GBT
beams at successively higher frequencies \citep{mhl+02}.  We used a
number of dense observations early in the campaigns to characterize
the orbits of binary pulsars.  The majority of timing observations
were carried out at $820\; \MHz$, but most pulsars were also observed
at other frequencies, allowing us to explore their spectral
properties.  We also started using the new Green Bank Ultimate Pulsar
Processing
Instrument\footnote{\url{https://safe.nrao.edu/wiki/bin/view/CICADA/NGNPP}}
(GUPPI) \citep{drd+08} in 2008 October.  Compared to the Spigot
back-end, GUPPI offers larger bandwidth, better frequency and time
resolution, higher dynamic range, and greater resilience to strong
RFI.

\subsection{Pulsar Timing Analysis \label{sec:timing}}

All of the new pulsars were observed regularly with the GBT as part of
our timing campaign.  Long period pulsars (with $P > 0.1\; \s$) were
observed for a minimum of about 11 months, while MSPs and recycled
pulsars were observed for a minimum of 20.  Each pulsar was typically
observed for $10$--$15\; \min$ per observing session.  High-\sn\
average pulse profiles were created for each observing frequency by
summing data from multiple observations.  We created standard pulse
profiles by fitting one or more Gaussians to these average pulse
profiles using a least squares minimization
routine\footnote{\texttt{pygaussfit.py} in \texttt{PRESTO}}.  These
standard profiles were used to compute pulse times of arrival (TOAs)
using either \texttt{PRESTO} or
\texttt{PSRCHIVE}\footnote{\url{http://psrchive.sourceforge.net/}}
\citep{hvm04} (depending on data format) by cross-correlation in the
Fourier domain.  We typically obtained two TOAs per observation for
isolated pulsars and four to six TOAs per observation for binary
pulsars, ensuring good sampling of the orbit.  Phase connected timing
solutions were created using the
\texttt{TEMPO}\footnote{\url{http://tempo.sourceforge.net}} software
package and the DE405 Solar System ephemeris.  All of our timing
solutions are referenced to UTC(NIST).  All the pulsars timed here
have timing solutions with reduced $\chi^2 > 1$.  Since we observe no
unmodeled trends in our timing residuals, this is probably due to an
underestimate of individual TOA uncertainties.  As is standard
practice, we multiplied all TOA uncertainties by an ``error factor'',
such that the reduced $\chi^2 = 1$.

\subsection{Flux Measurements \label{sec:flux}}

Mean flux densities were estimated by assuming that the off-pulse
root-mean-square (RMS) noise level was described by the radiometer
equation,
\begin{eqnarray}
  \sigma = \frac{\beta T\rmsub{sys}}
  {G \sqrt{n\rmsub{p} \Delta \nu\, t\rmsub{int}}},
\end{eqnarray}
where $T\rmsub{sys}$ is the total system temperature.  To ensure a
proper estimate of the RMS noise level, we fit a third order
polynomial to the off-pulse region and then subtracted this to create
a flat off-pulse baseline.  It is important to keep in mind, however,
that observed pulsar fluxes are variable due to interstellar
scintillation.  The values that we report here are determined by
averaging several observations but should be treated only as
representative.  We also calculated the spectral index when flux
density estimates were available for multiple frequencies.  We did
this by fitting a standard power law to the flux density estimates,
typically at $350\; \MHz$ and $820\; \MHz$, assuming $S_{\nu} \propto
\nu^{\alpha}$.  The average for the pulsars presented here is $\langle
\alpha \rangle = -1.7$, which is very similar to the average value
presented in \citet{lylg95}.

We also attempted to measure the rotation measure (RM) whenever fully
calibrated polarization data were available (which was at least once
for each pulsar).  We searched over a wide range of RMs, from $\pm
1000\; \mathrm{rad\; \pmsq}$.  We could only detect a significant RM
for a subset of pulsars.  Those pulsars without reported RMs are
probably weakly polarized sources.

\section{Results \label{sec:results}}

\begin{figure}
\centering
\includegraphics[width=5.75in]{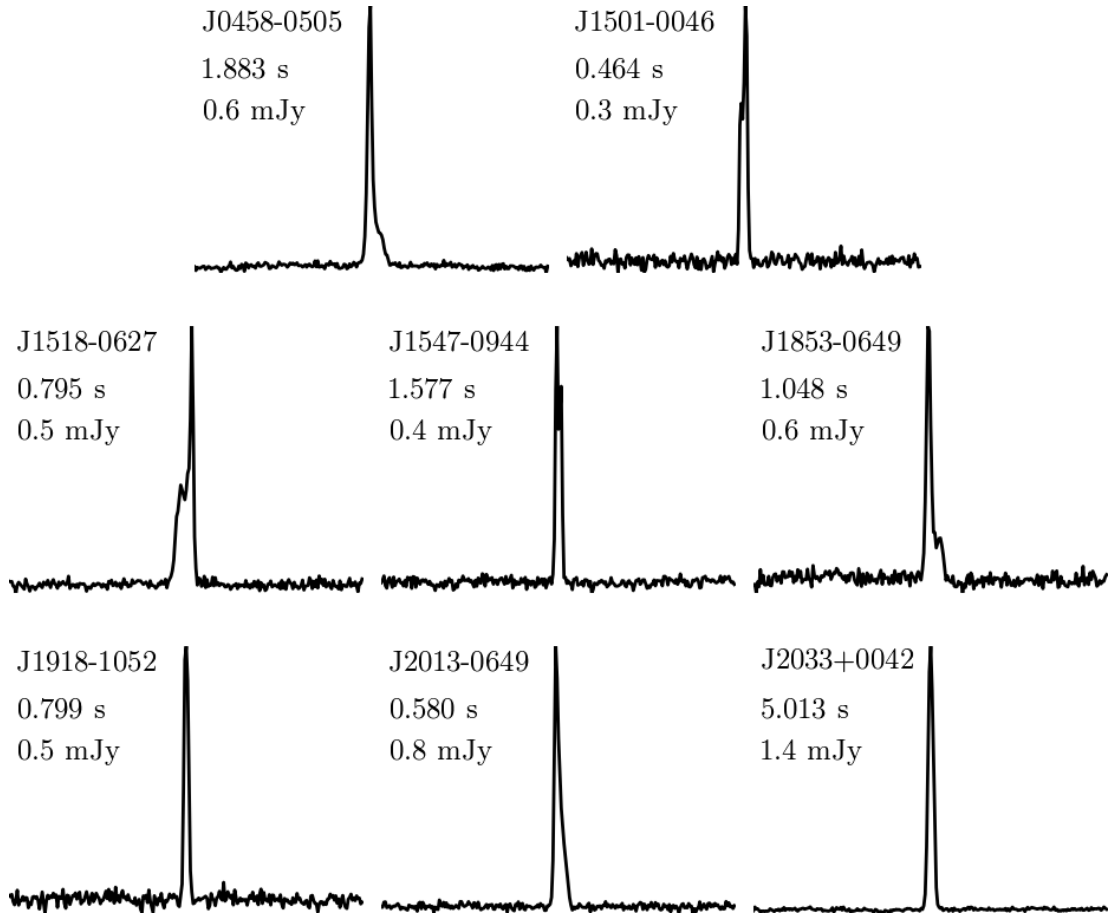}
\caption{Integrated $820\; \MHz$ pulse profiles for the newly
  discovered long period pulsars presented here.  All profiles show
  one full rotation of the pulsar (i.e., from phase 0--1) with 256
  phase bins.  The profiles were made by adding all the RFI-free
  observations for each pulsar and were used to create standard pulse
  profiles at $820\; \MHz$.  Pulse periods and mean flux densities are
  also given. \label{fig:slow_profs}}
\end{figure}

\begin{figure}
\centering
\includegraphics[width=5.7in]{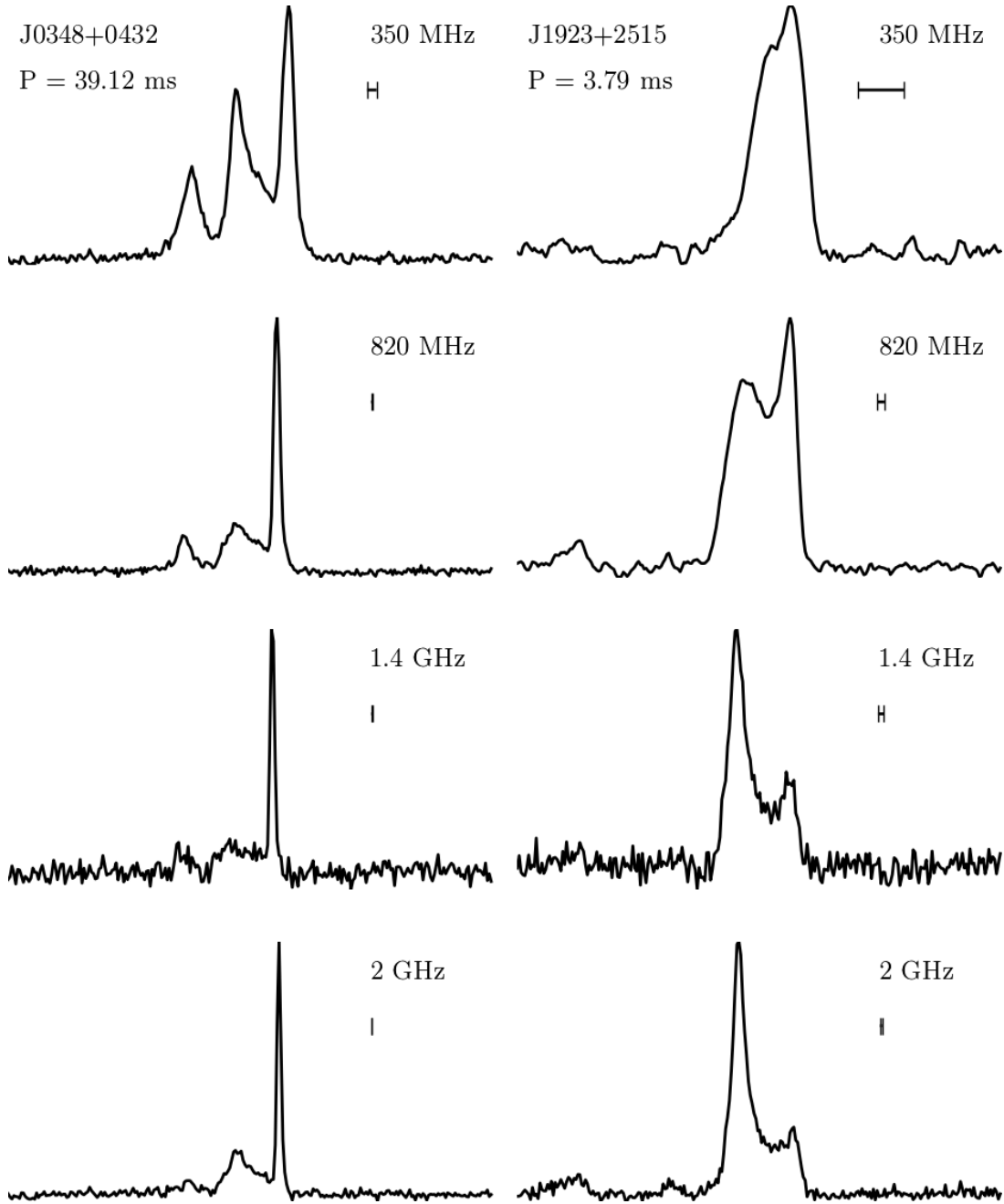}
\caption{Integrated pulse profiles at four observing frequencies for
  the newly discovered recycled pulsars presented here.  All profiles
  show one full rotation of the pulsar (i.e., from phase 0--1) with
  256 phase bins.  The summed profiles were made by aligning each
  folded profile using the \texttt{TEMPO} ephemeris and then adding
  all the RFI-free observations at the specified frequencies.  The
  profile evolution in both pulsars is clear.  The bars show the
  relative timescale for dispersive smearing at each frequency (see
  Equation \ref{eqn:dm_smear}).  \label{fig:recycled_profs}}
\end{figure}

\begin{figure}
\centering
\includegraphics[width=5.75in]{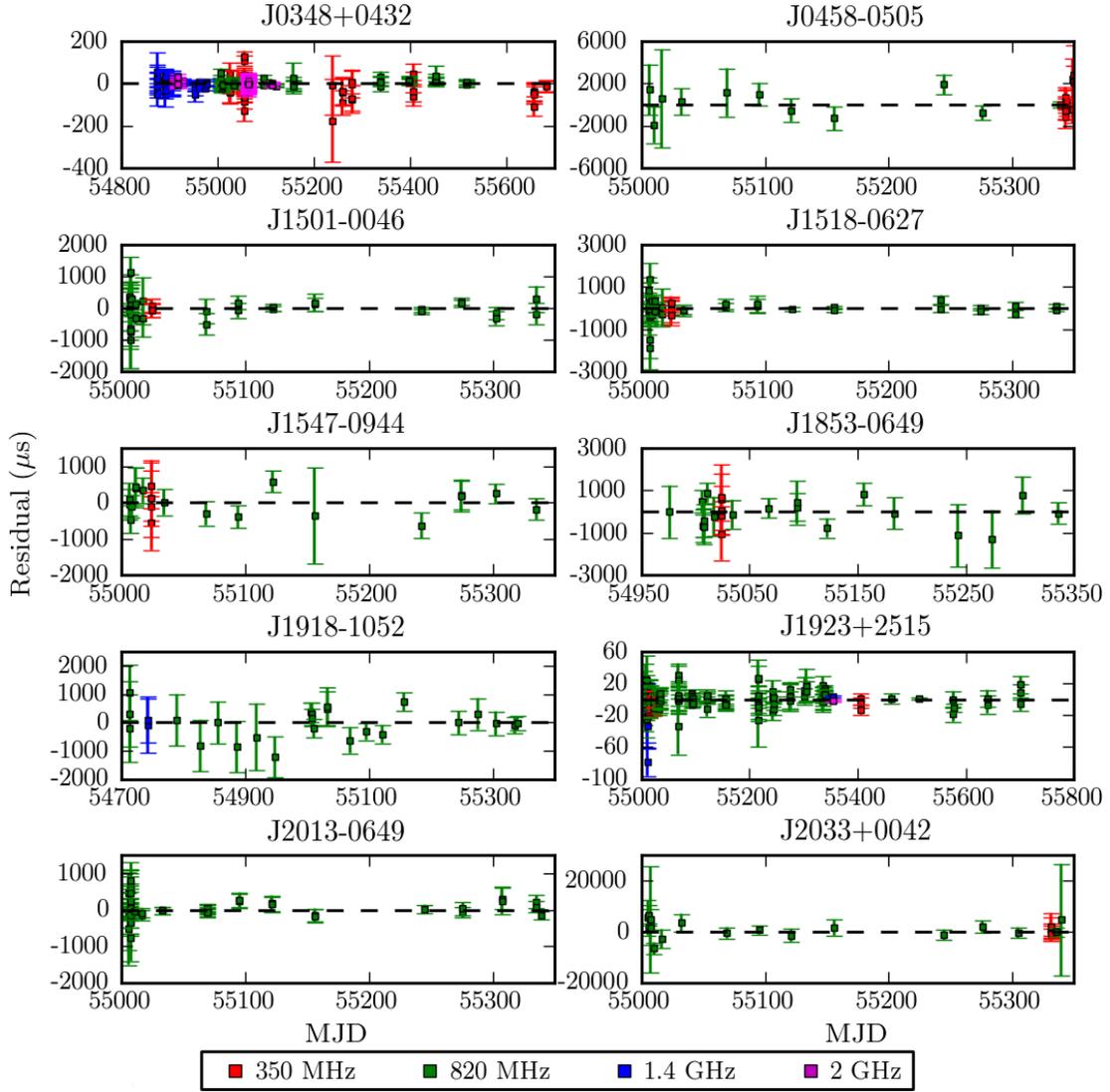}
\caption{Post-fit timing residuals for each of the newly discovered
  pulsars.  Only phase connected TOAs are shown.  Nominal TOA errors
  have been multiplied by a constant error factor so that the reduced
  $\chi^2 = 1$.  Note that the axes have different scales in most
  plots. \label{fig:residuals}}
\end{figure}

A total of 31 new pulsars have been discovered thus far in the
Drift-scan Survey.  The first 13 are presented in \citetalias{dss} and
10 are discussed here.  As mentioned in \S\ref{sec:intro}, PSR
J1023$+$0038 has been discussed elsewhere by \citep{asr+08} and
\citep{,akb+10}, while another MSP, PSR J2256$-$1024, will be
presented in a future paper (Stairs et al. in prep).  Full timing
solutions have not been obtained for the six most recently discovered
pulsars and these will be presented in future work.  The 10 pulsars
presented here include eight long period, isolated pulsars, one mildly
recycled binary pulsar, and one isolated MSP.  Seven of these 10
pulsars were detected independently in our searches for single pulses.
Full timing solutions and other properties for the long-period pulsars
are presented in Table \ref{table:slow_params} and for the recycled
pulsars in Table \ref{table:recycled_params}.  Integrated pulse
profiles can be seen in Figures \ref{fig:slow_profs} and
\ref{fig:recycled_profs} and post-fit timing residuals are shown in
Figure \ref{fig:residuals}.  We discuss some individual systems in
greater detail below.

\subsection{PSR J0348$+$0432: A Relativistic Binary with a Low-Mass
  Companion \label{sec:J0348}}

PSR J0348$+$0432 (hereafter J0348) is a mildly recycled binary pulsar
with $P = 39.1\; \ms$.  The low magnetic field of J0348
($B\rmsub{surf} = 3.1 \times 10^9\; \gauss$) indicates that it is
indeed partially recycled and not a young pulsar\footnote{Although we
  have not measured proper motion and cannot calculate its
  contribution to $\dot{P}\rmsub{obs}$, it is certainly not sufficient
  to increase the $B\rmsub{surf}$ by several orders of magnitude.}.
The DM of J0348 is $40.5\; \dmu$ and the DM-derived distance is $2.1\;
\kpc$.  The orbital period of this system is $2.4\; \hr$, and only
three pulsars with $P < 40\; \ms$ outside of globular clusters have
shorter periods.  If we assume a mass of $1.4\; \Msun$ for J0348 then
the observed mass function, $f(M) = 2.9 \times 10^{-4}$, implies a
minimum companion mass of $0.086\; \Msun$.  We searched for and
identified an optical counterpart to J0348 in the Sloan Digital Sky
Survey with corrected SDSS magnitudes $u' = 21.84 \pm 0.19$, $g' =
20.71 \pm 0.03$, $r' = 20.60 \pm 0.03$, $i' = 20.69 \pm 0.05$, and $z'
= 20.40 \pm 0.15$.  More detailed spectroscopic follow-up with the
Apache Point $3.5$-$\m$ telescope and the Very Large Telescope have
shown that the companion to J0348 is a low-mass white dwarf.  The
combination of a neutron star and low-mass white dwarf in a tight,
relativistic orbit is unique among pulsars and makes J0348 an
excellent laboratory for testing general relativity and other theories
of gravity.  Specifically, theories that invoke a scalar gravitational
field predict that J0348 will be a strong emitter of dipolar
gravitational radiation because the very different binding energies of
the neutron star and white dwarf will lead them to couple differently
to the scalar field \citep[e.g.][]{sta03}.  Similar tests have been
done with PSR J1141$-$6545 \citep{bbv08}, J1012$+$5307 \citep{lwj+09}
and J1738$+$0333 \citep{fwe+12}, but J0348 is in a tighter, more
relativistic orbit and likely has a less massive companion, so it will
be a stronger probe of these effects.  A full analysis of the
spectroscopic observations of J0348 and their implications for
alternative theories of gravity will be presented in Antoniadis et al.
(in prep).

J0348 shows significant profile evolution as a function of frequency
(see Figure \ref{fig:recycled_profs}); at frequencies above $1.4\;
\GHz$, the main profile component becomes extremely narrow, with a
duty cycle of $\sim$$1\%$.  This has allowed us to obtain very precise
pulse arrival times---our RMS timing residuals are $9.3\; \us$ but at
high frequencies individual TOA uncertainties can be $\lesssim 3\;
\us$.  We are continuing long-term timing of this pulsar using the
Arecibo Observatory.

\subsection{PSR J1923$+$2515 \label{sec:J1923}}

PSR J1923$+$2515 (hereafter J1923) is an isolated MSP with a
$3.8$-$\ms$ spin period.  Figure \ref{fig:recycled_profs} shows the
integrated pulse profile of J1923 at several different frequencies and
the evolution in the profile shape is clear.  We see evidence for a
weak interpulse in our summed $820\; \MHz$ and $2\; \GHz$ data.  The
timing of J1923 improved significantly at higher frequencies, where we
were able to obtain TOAs with uncertainties $\lesssim 1\; \us$.  J1923
is being regularly observed at Arecibo as part of the NANOGrav timing
array for gravitational wave detection \citep{dfg+12}.  It will also
be suitable for the European Pulsar Timing Array \citep{vlj+11}.

J1923 is the only pulsar presented here for which we were able to
measure a significant proper motion.  We find $\mu_{\alpha} =
-6.2(1.2)\; \mathrm{mas}\, \pyr$ and $\mu_{\delta} = -23.6(3.5)\;
\mathrm{mas}\, \pyr$.  We performed an F-test to determine if the
addition of proper motion is in fact required by the data.  The full
$\chi^2$ of our timing model excluding proper motion is 339.63, with
147 degrees of freedom.  When proper motion is included in the fit,
$\chi^2 = 146.02$ with 145 degrees of freedom.  The probability that
this improvement is due to chance is $1 \times 10^{-16}$.  Thus, the
improvement in our timing solution when proper motion is included is
extremely significant.  We used the DM and the NE2001 model of
Galactic free electron density to estimate the distance to J1923, $D =
1.6(3)\; \kpc$, where the number in parentheses represents a 20\%
fractional error, which is typical for these estimates \citep{cl02}.
At this distance, the observed proper motion corresponds to a
transverse velocity $v_{\perp} = 188(46)\; \km\, \ps$, which is within
the observed range of other MSPs, though higher than average
\citep{gg00,bpl+02,gsf+11}.

Using this $v_{\perp}$ we can calculate the magnitude of the Shklovskii
effect \citep{shk70}:\
\begin{eqnarray}
\frac{\dot{P}_{\mu}}{P} = \frac{v_{\perp}^2}{c\; D}.
\end{eqnarray}
We find $\dot{P}_{\mu} = 8.9(4.7) \times 10^{-21}\; \s\, \ps$.
Acceleration within the Galactic potential will also cause a bias in
the observed $\dot{P}$.  We estimate this contribution following
\citet{dt95}, but find that biases due to acceleration perpendicular
and parallel to the Galactic plane are only $-3.3 \times 10^{-22}\;
\s\, \ps$ and $-3.4 \times 10^{-22}\; \s\, \ps$, respectively.  These
are an order of magnitude smaller than $\dot{P}_{\mu}$.  The bias due
to the Shklovskii effect is 94\% of the observed $\dot{P}$ and would
imply that the intrinsic spin-down of the pulsar is
$\dot{P}\rmsub{int} = 0.6(4.7) \times 10^{-21}\; \s\, \ps$, so we can
only place an upper limit on $\dot{P}\rmsub{int}$ at this time.  A
more precise measurement of proper motion or a better distance
estimate will be needed to constrain the magnitude of the Shklovskii
effect and to obtain a better measurement of $\dot{P}\rmsub{int}$.  In
the meantime, the derived quantities listed in Table
\ref{table:recycled_params} are upper limits based upon our
measurement uncertainties for $\dot{P}\rmsub{int}$.

\subsection{PSR J0458$-$0505 \label{sec:J0458}}

PSR J0458$-$0505 (hereafter J0458) is a nulling pulsar with a
$1.9$-$\s$ spin period.  It was detected in both the Fourier domain
and single-pulse searches.  The profile is slightly asymmetric (see
Figure \ref{fig:slow_profs}), with a small trailing component.  The
reduced $\chi^2$ obtained from our timing solution after fitting for
position, $P$, $\dot{P}$, and DM was substantially higher than unity.
As we see no systematic trends in our timing residuals, we assume that
the number of pulses per individual observation was too small, due to
a combination of a long period, limited integration time and a very
large nulling fraction (see below). The pulse profile thus probably
did not stabilize within these TOA integrations.  To make the reduced
$\chi^2$ equal to one, we multiplied our individual TOA errors by a
constant factor of $2.97$.  Although we were still able to derive an
accurate timing solution for J0458, the fractional errors in the
timing parameters are larger than for most of the other pulsars
presented here, especially for declination and $\dot{P}$.

\subsubsection{Estimate of the Nulling Fraction \label{sec:null}}

We estimate the nulling fraction, NF, of J0458 in the following way.
We first removed strong sources of RFI from each observation.  We then
folded each data-set using sub-integrations that were a single pulse
period in duration using the \texttt{psrfits\_singlepulse} command
from
\texttt{psrfits\_utils}\footnote{\url{https://github.com/demorest/psrfits_utils}}.
In some sub-integrations, systematic trends due to lower levels of RFI
were still visible.  To remove these, we used a least-squares
minimization technique to fit up to a maximum of four independent
sinusoids to the off-pulse region of each sub-integration and then
subtracted them from the data, creating a flat off-pulse baseline.
Each sub-integration was then normalized to have an off-pulse median
and RMS noise level of zero and one, respectively.

\begin{figure}
\centering
\begin{tabular}{c}
\includegraphics[height=3.5in]{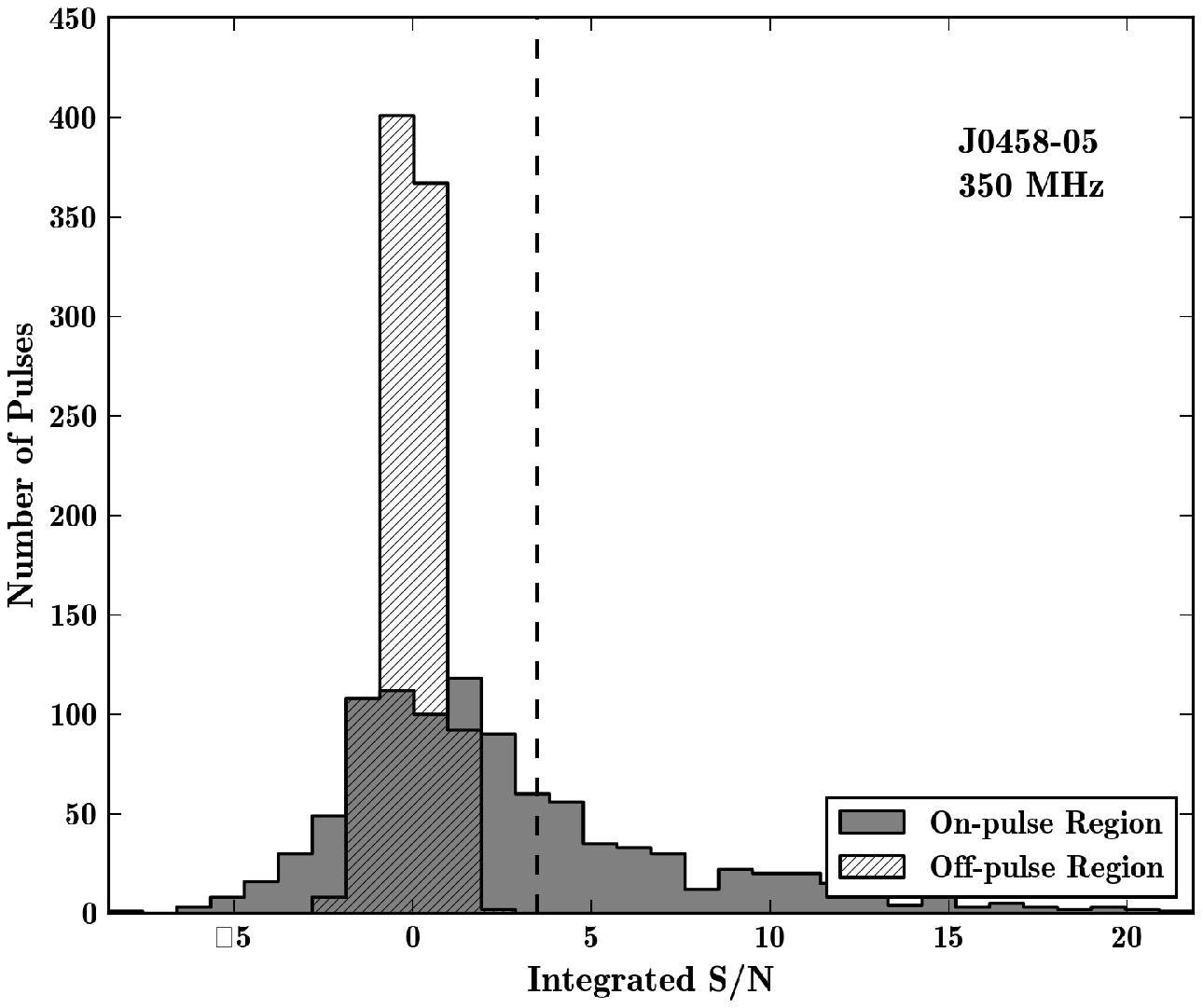} \\ 
\includegraphics[height=3.5in]{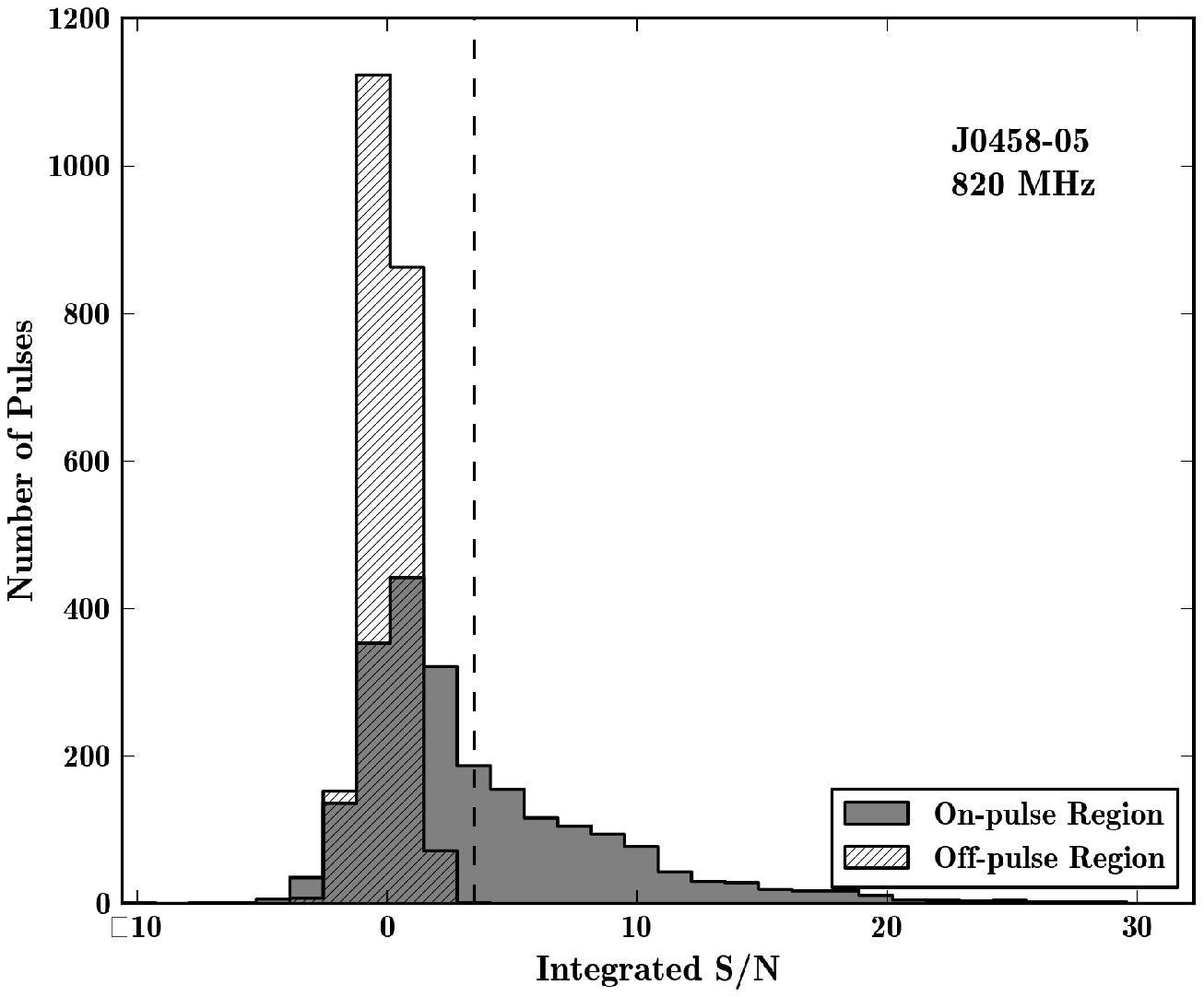} \\
\end{tabular}
\caption{Histograms of integrated \sn\ for single pulses of J0458.
  The top panel is for data taken at $350\; \MHz$, while the bottom
  panel is for data taken at $820\; \MHz$.  The on-pulse region is
  shown in gray and the off-pulse region as hatched.  The dashed line
  shows the $\sn\rmsub{thresh} = 3.5$, above which we counted the
  pulsar as being in an ``on'' state.
  \label{fig:J0458_snr_hist}}
\end{figure}

\begin{figure}
\centering
\begin{tabular}{c}
\includegraphics[height=3.5in]{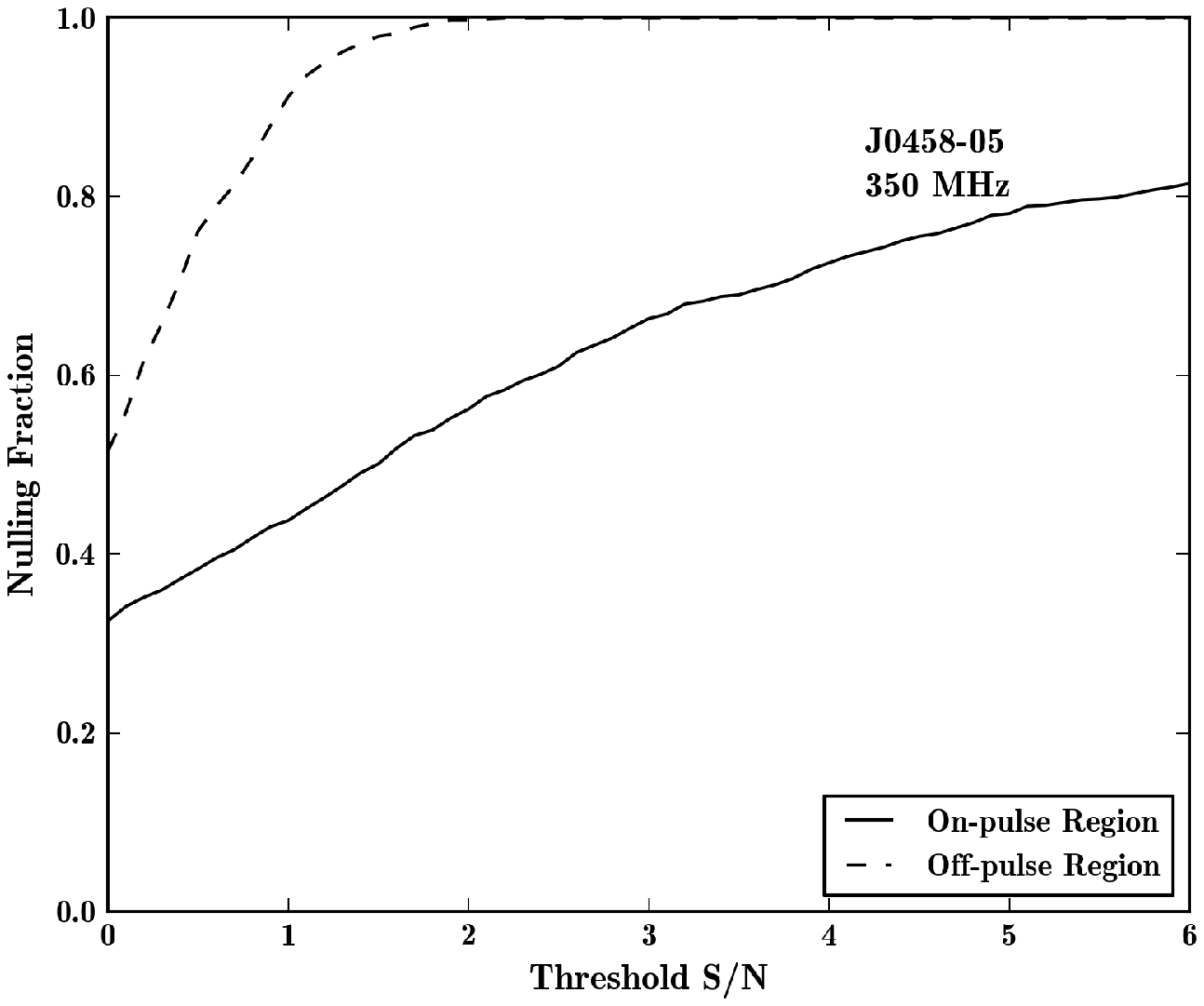} \\ 
\includegraphics[height=3.5in]{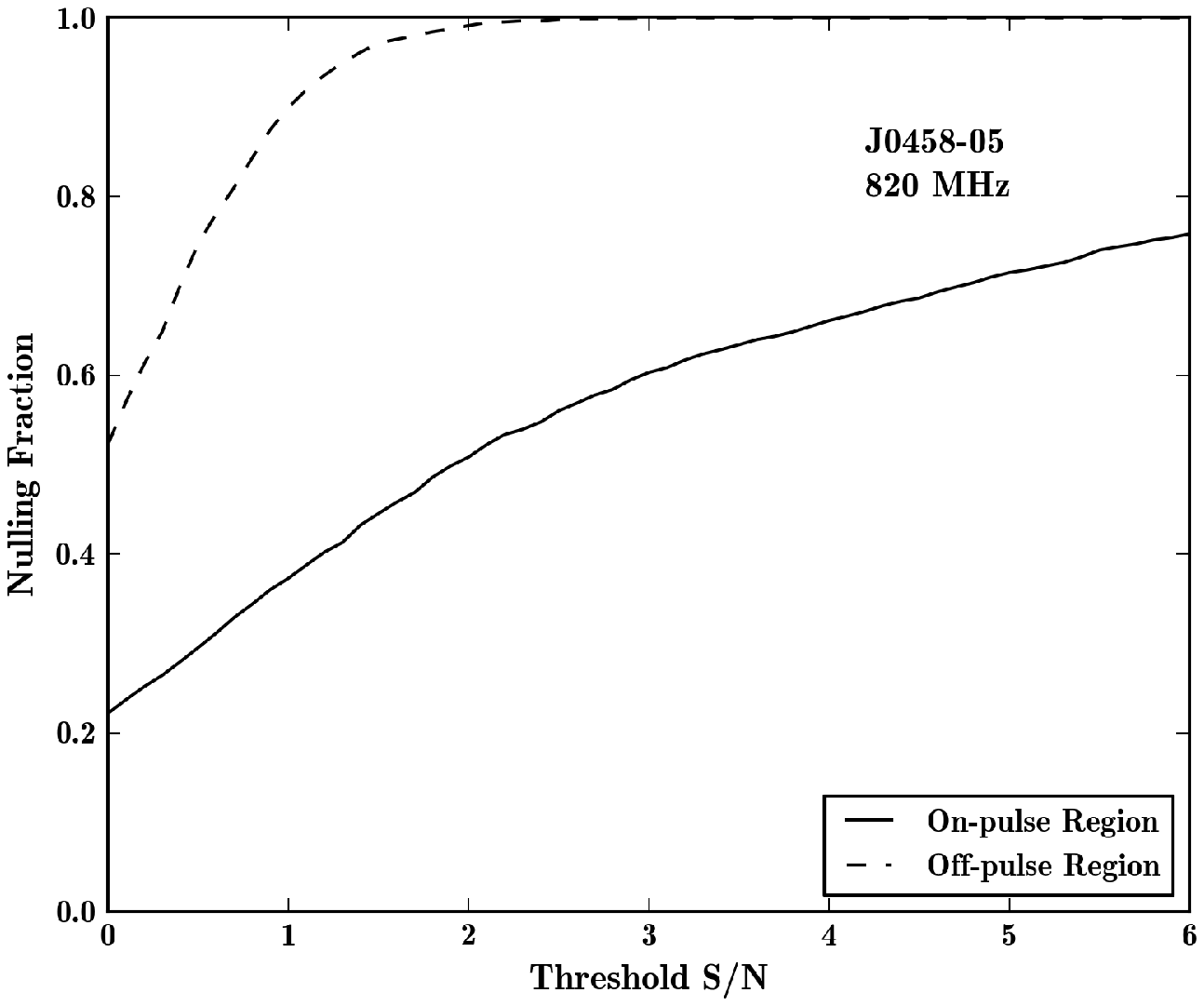} \\
\end{tabular}
\caption{The nulling fraction of J0458 as a function of
  $\sn\rmsub{thresh}$.  The top panel is for data taken at $350\;
  \MHz$, while the bottom panel is for data taken at $820\; \MHz$.
  \label{fig:J0458_snr_nfs}}
\end{figure}

To determine if the pulsar was in an ``on'' state, we calculated the
integrated \sn\ in the on-pulse region, which was determined by
inspection of the integrated pulse profile.  We also calculated the
integrated \sn\ in an off-pulse region with the same number of bins as
the on-pulse region.  We counted a pulse as being in the ``on'' state
if it had an integrated \sn\ above some threshold,
$\sn\rmsub{thresh}$.  We chose $\sn\rmsub{thresh}$ based on the
statistics of the off-pulse region.  Histograms of the on-pulse and
off-pulse S/Ns can be found in Figure \ref{fig:J0458_snr_hist} and in
Figure \ref{fig:J0458_snr_nfs} we show NF as a function of threshold
\sn.  Our calculations show that the off-pulse region rarely exceeded
$\sn = 3.0$, as expected for Gaussian distributed noise.  To be
conservative, we set $\sn\rmsub{thresh} = 3.5$, though we also report
NF for $\sn\rmsub{thresh} = 3.0$ and $4.0$ for comparison.

After removing RFI, J0458 was observed for a total of 2218 full
rotations at $820\; \MHz$ and we find $\mathrm{NF} = [0.60,0.63,0.66]$
for $\sn\rmsub{thresh} = [3.0,3.5,4.0]$, respectively.  J0458 was
observed for a total of 978 rotations at $350\; \MHz$ with
$\mathrm{NF} = [0.66,0.69,0.73]$ for $\sn\rmsub{thresh} =
[3.0,3.5,4.0]$, respectively.  These $\mathrm{NFs}$ are fairly high
compared to other pulsars, but are not unprecedented --- PSRs B1112+50
and B1944+17 have comparable $\mathrm{NF}$ \citep{rit76}.  The nulling
fraction of J0458 seems to be similar at both $350\; \MHz$ and $820\;
\MHz$.  This behavior is consistent with previous studies that suggest
nulling is a broadband phenomenon at low frequencies \citep[see][and
references therein]{big92}.

\subsection{PSR J2033$+$0042 \label{sec:J2033}}

PSR J2033$+$0042 (hereafter J2033) is a $5.0$-$\s$ pulsar that, like
J0458, nulls significantly.  It was reported by \citet{bb10} as a RRAT
that sometimes was detected in Fourier searches.  \citet{bb10} report
on the position, period, and DM of J2033 and note its high nulling
fraction and the presence of drifting sub-pulses.  Here, we present a
full timing solution and quantitative measurement of the nulling
fraction.  We detected J2033 in both single-pulse and Fourier
searches.  Only nine radio pulsars and 12 RRATs in the ATNF catalog
have a longer period than J2033.  Like J0458, the fractional errors in
our timing parameters were relatively large, probably because our
integration times were shorter than the pulse stabilization time.
J2033 has a longer period than J0458 and also nulls significantly.

\begin{figure}
\centering
\begin{tabular}{c}
\includegraphics[height=3.5in]{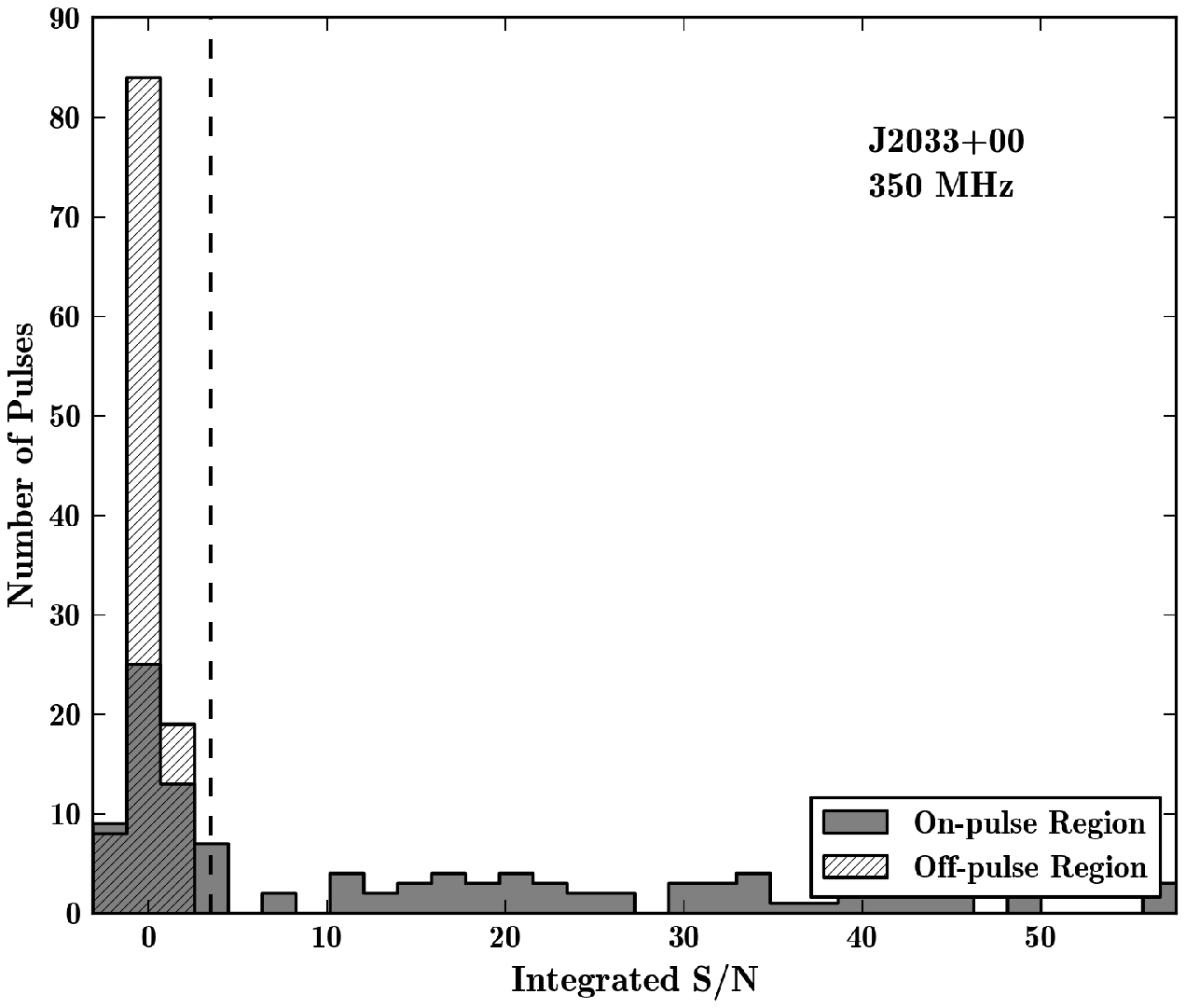} \\ 
\includegraphics[height=3.5in]{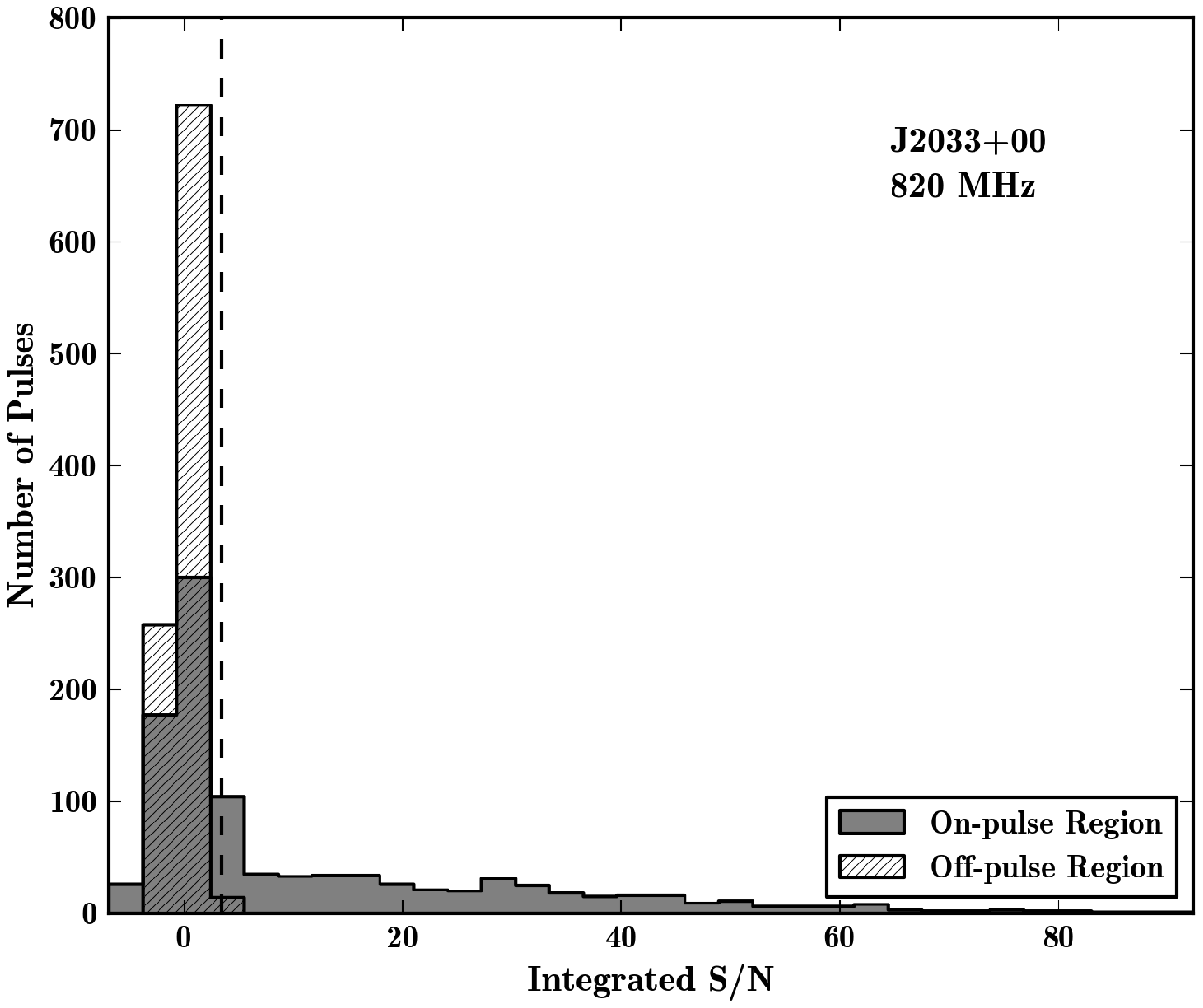} \\
\end{tabular}
\caption{Histograms of \sn\ for single pulses of J2033.  The data
  labels are the same as in Figure \ref{fig:J0458_snr_hist}.
  \label{fig:J2033_snr_hist}}
\end{figure}

\begin{figure}
\centering
\begin{tabular}{c}
\includegraphics[height=3.5in]{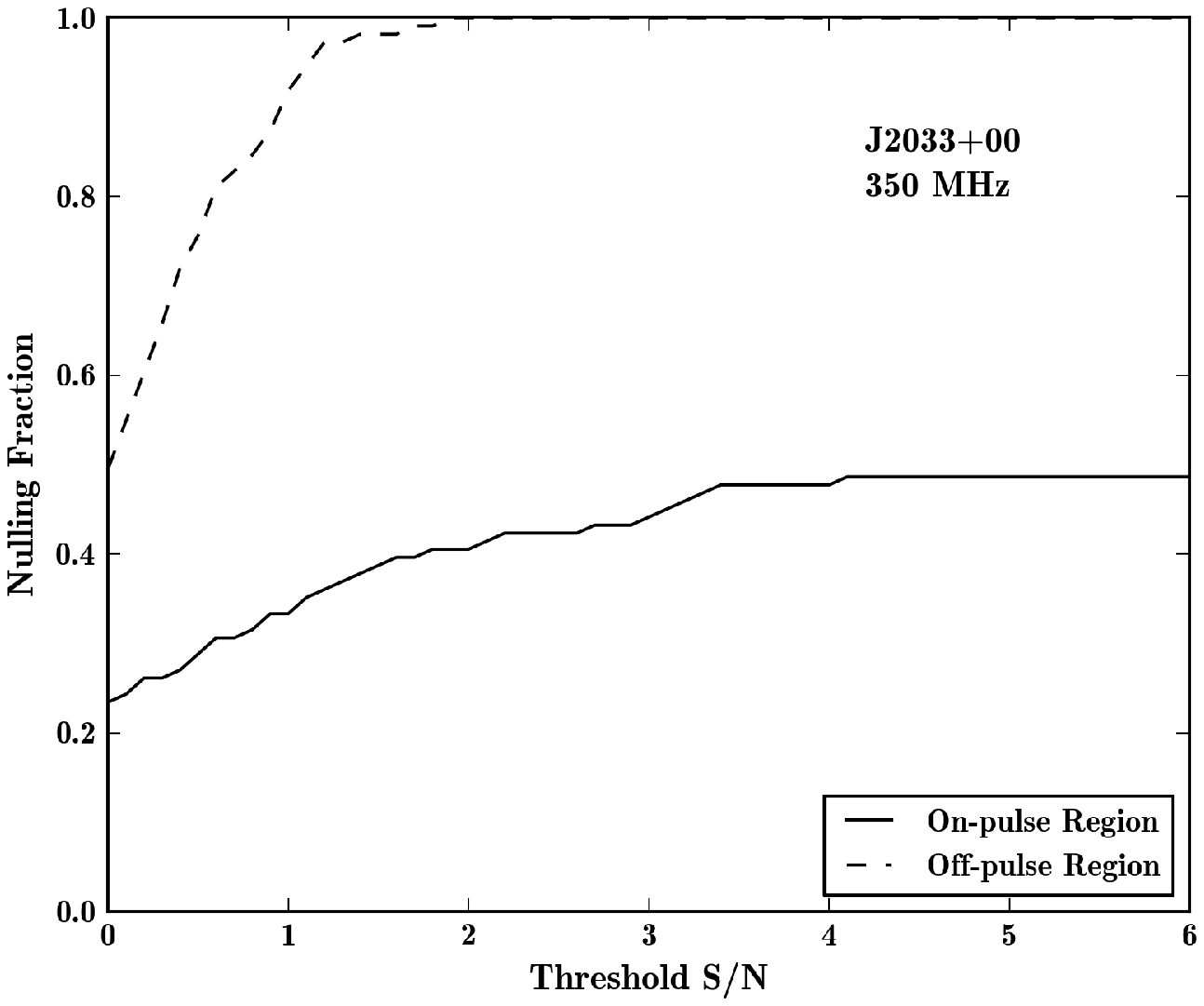} \\ 
\includegraphics[height=3.5in]{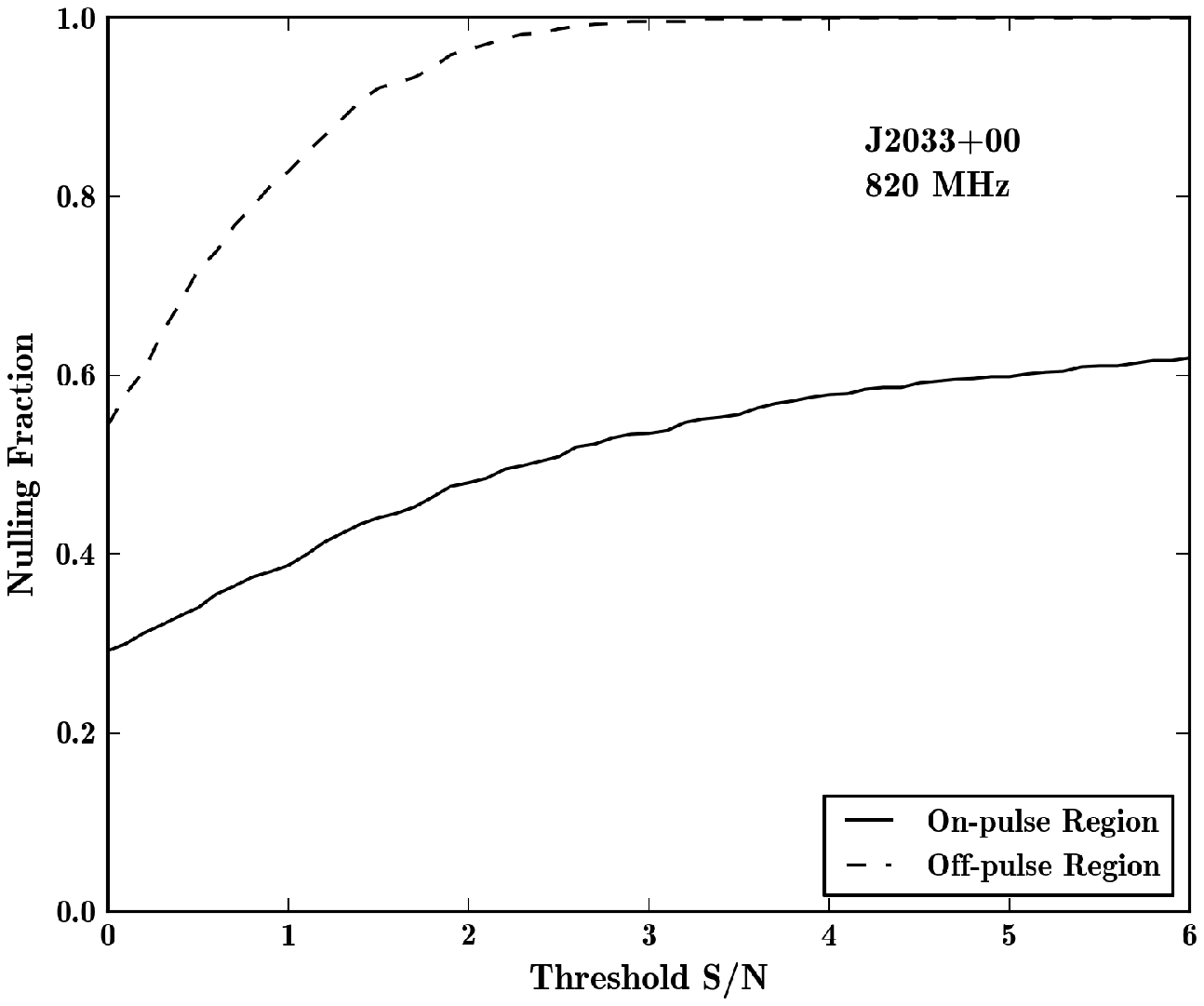} \\
\end{tabular}
\caption{The nulling fraction of J2033 as a function of
  $\sn\rmsub{thresh}$.  The data labels are the same as in Figure
  \ref{fig:J0458_snr_nfs}. \label{fig:J2033_snr_nfs}}
\end{figure}

We used the same procedure as outlined in \S\ref{sec:null} to estimate
the nulling fraction in J2033.  We observed the pulsar for 994
rotations at $820\; \MHz$ and we find a $\mathrm{NF} =
[0.53,0.56,0.58]$ for $\sn\rmsub{thresh} = [3.0,3.5,4.0]$,
respectively.  J2033 was observed for 111 rotations at $350\; \MHz$
and we find $\mathrm{NF} = [0.44,0.48,0.49]$ for $\sn\rmsub{thresh} =
[3.0,3.5,4.0]$, respectively.  Histograms of \sn\ can be found in
Figure \ref{fig:J2033_snr_hist} and $\mathrm{NF}$ as a function of
$\sn\rmsub{thresh}$ is plotted in Figure \ref{fig:J2033_snr_nfs}.  The
nulling fraction of J2033 is somewhat higher at $820\; \MHz$ than at
$350\; \MHz$, but we only observed J2033 at $350\; \MHz$ on one
occasion, so a more detailed study of the nulling characteristics of
this pulsar should be conducted before drawing firm conclusions about
the frequency dependence of $\mathrm{NF}$.  Overall, pulsars J0458 and
J2033 are on less than half of the time, adding to a growing
sub-population of pulsars that are mostly off \citep{kkl+11}.

\section{Conclusion \label{sec:conc}}

The Drift-scan Survey has discovered 31 pulsars, of which 10 are
presented here.  The majority are isolated long-period pulsars.  J0348
is a mildly recycled binary pulsar that has a low-mass white dwarf
companion in a relativistic orbit. It is a unique and powerful system
for testing gravitational theories and hence we are continuing to time
it long-term.  A more detailed study of J0348 will be presented in
Antoniadis et al. (in prep).  J1923 is an isolated MSP.  We have a
significant measurement of the pulsar's proper motion, but the implied
magnitude of the Shklovskii effect is nearly equal to the observed
spin-down, so we are only able to set limits on the rotational
characteristics of J1923.  Long-term monitoring should help to better
constrain the proper motion and intrinsic spin-down.  J0458 and J2033
are both nulling pulsars with nulling fractions $\gtrsim 50\%$.

R.S.L.\ was a student at the National Radio Astronomy Observatory and
was supported through the GBT Student Support program and the National
Science Foundation grant AST-0907967 during the course of this work.
D.R.L., M.A.M., and J.B.\ acknowledge support from a WVEPSCoR Research
Challenge Grant.  J.W.T.H.\ is a Veni Fellow of the Netherlands
Foundation for Scientific Research.  Pulsar research at UBC is
supported by an NSERC Discovery Grant and Special Research Opportunity
grant as well as the Canada Foundation for Innovation.  V.M.K.\ holds
the Lorne Trottier Chair in Astrophysics and Cosmology, and a Canada
Research Chair, a Killam Research Fellowship, and acknowledges
additional support from an NSERC Discovery Grant, from FQRNT via le
Centre de Recherche Astrophysique du Queb\'{e}c and the Canadian
Institute for Advanced Research.  R.F.C., C.E.R., and T.P.\ were
summer students at the National Radio Astronomy Observatory during a
portion of this work.  We thank Paulo Freire for refereeing this
manuscript and providing helpful feedback.  We are also grateful to
NRAO for a grant that assisted data storage.  The National Radio
Astronomy Observatory is a facility of the National Science Foundation
operated under cooperative agreement by Associated Universities, Inc.


\begin{deluxetable}{lc}
  \centering
  \tabletypesize{\small}
  \tablewidth{0pt}
  \tablecolumns{2}
  \tablecaption{Parameters of the GBT $350\; \MHz$ Drift-scan
    Survey \label{table:surv_params}}
  \tablehead{
    \colhead{Paramter}          &
    \colhead{Value}}
  \startdata
    ADC Conversion Factor, $\beta$                   & 1.16                  \\
    Signal-to-noise threshhold, $\sn\rmsub{min}$     & 6.0                   \\
    Receiver Temperature, $T\rmsub{rec}$ ($\K$)      & 46                    \\
    Telescope Gain, $G$ ($\K\; \Jy^{-1}$)            & 2.0                   \\
    Number of summed polarizations, $n\rmsub{p}$     & 2                     \\
    Length of pseudo-pointing, $t\rmsub{int}$ ($\s$) & 140                   \\
    Bandwidth, $\Delta f$ ($\MHz$)                   & 50                    \\
    Number of frequency channels, $n\rmsub{chan}$    & 2048\tablenotemark{a} \\
    Sampling time, $t\rmsub{samp}$ ($\us$)           & 81.92                 \\
  \enddata
  \tablenotetext{a}{A small amount of data was recorded with 1024
    frequency channels early in the survey.}
\end{deluxetable}

\begin{deluxetable}{lccc}
  \centering
  \tabletypesize{\footnotesize}
  \tablewidth{0pt}
  \tablecolumns{4}
  \tablecaption{Parameters of Newly Discovered Long Period Pulsars \label{table:slow_params}}
  \tablehead{
    \colhead{Parameter} &
    \colhead{PSR J0458$-$0505} &
    \colhead{PSR J1501$-$0046} &
    \colhead{PSR J1518$-$0627}}
  \startdata
    \cutinhead{Timing Parameters}
    Right Ascension (J2000)                       \dotfill & 04:58:37.121(26)              & 15:01:44.9558(94)             & 15:18:59.1104(80)             \\
    Declination (J2000)                           \dotfill & $-$05:05:05.1(4.0)            & $-$00:46:23.52(88)            & $-$06:27:07.70(66)            \\
    Pulsar Period ($\s$)                          \dotfill & 1.88347965849(18 )            & 0.4640368139284(82)           & 0.7949966745699(78)           \\
    Period Derivative ($\s\, \ps$)                \dotfill & 5.3(1.5)$\times$10$^{-16}$    & 2.391(60)$\times$10$^{-16}$   & 4.179(56)$\times$10$^{-16}$   \\
    Dispersion Measure (\dmu)                     \dotfill & 47.806(32)                    & 22.2584(90)                   & 27.9631(98)                   \\
    Reference Epoch (MJD)                         \dotfill & 55178.0                       & 55170.0                       & 55170.0                       \\
    Span of Timing Data (MJD)                     \dotfill & 55006--55349                  & 55006--55335                  & 55006--55335                  \\
    Number of TOAs                                \dotfill & 22                            & 39                            & 43                            \\
    RMS Residual ($\us$)                          \dotfill & 716                           & 172                           & 175                           \\
    Error Factor                                  \dotfill & 2.967                         & 1.095                         & 1.050                         \\
    \cutinhead{Derived Parameters}
    Galactic Longitude (deg)                      \dotfill & 204.14                        & 356.58                        & 355.15                        \\
    Galactic Latitude (deg)                       \dotfill & $-$27.35                      & 48.05                         & 41.0                          \\
    DM Derived Distance (kpc)                     \dotfill & 2.6                           & 1.4                           & 1.6                           \\
    Surface Magnetic Field ($10^{12}$ Gauss)      \dotfill & 1.0                           & 0.33                          & 0.58                          \\
    Spin-down Luminosity ($10^{32}\; \erg\, \ps$) \dotfill & 0.032                         & 0.95                          & 0.33                          \\
    Characteristic Age (Myr)                      \dotfill & 56                            & 31                            & 30                            \\
    $820\; \MHz$ FWHM                             \dotfill & 0.014                         & 0.022                         & 0.012                         \\
    $820\; \MHz$ Flux Density ($\mJy$)            \dotfill & 0.5                           & 0.3                           & 0.4                           \\
    Spectral Index                                \dotfill & $-1.6$                        & $-2.1$                        & $-1.8$                        \\
  \enddata
\end{deluxetable}
\setcounter{table}{1}
\begin{deluxetable}{lccc}
  \centering
  \tabletypesize{\footnotesize}
  \tablewidth{0pt}
  \tablecolumns{4}
  \tablecaption{Parameters of Newly Discovered Long Period Pulsars (Cont.)}
  \tablehead{
    \colhead{Parameter} &
    \colhead{PSR J1547$-$0944} &
    \colhead{PSR J1853$-$0649} &
    \colhead{PSR J1918$-$1052}}
  \startdata
    \cutinhead{Timing Parameters}
    Right Ascension (J2000)                       \dotfill & 15:47:46.058(36)              & 18:53:25.422(36)              & 19:18:48.247(13)             \\
    Declination (J2000)                           \dotfill & $-$09:44:7.8(3.2)             & $-$06:49:25.9(2.6)            & $-$10:52:46.38(66)            \\
    Pulsar Period ($\s$)                          \dotfill & 1.576924632943(44)            & 1.048132105087(54)            & 0.798692542358(15)           \\
    Period Derivative ($\s\, \ps$)                \dotfill & 2.938(36)$\times$10$^{-15}$   & 1.548(44)$\times$10$^{-15}$   & 8.653(15)$\times$10$^{-16}$  \\
    Dispersion Measure (\dmu)                     \dotfill & 37.416(22)                    & 44.541(36)                    & 62.73(80)                     \\
    Reference Epoch (MJD)                         \dotfill & 55170.0                       & 55170.0                       & 55026.0                       \\
    Span of Timing Data (MJD)                     \dotfill & 55006--55335                  & 54976--55335                  & 54712--55339                  \\
    Number of TOAs                                \dotfill & 22                            & 24                            & 26                            \\
    RMS Residual ($\us$)                          \dotfill & 338                           & 588                           & 394                           \\
    Error Factor                                  \dotfill & 1.630                         & 2.150                         & 2.480                         \\
    \cutinhead{Derived Parameters}
    Galactic Longitude (deg)                      \dotfill & 358.31                        & 27.08                         & 26.23                         \\
    Galactic Latitude (deg)                       \dotfill & 33.57                         & $-$3.55                       & $-$10.96                      \\
    DM Derived Distance (kpc)                     \dotfill & 1.9                           & 1.5                           & 2.1                           \\
    Surface Magnetic Field ($10^{12}$ Gauss)      \dotfill & 2.2                           & 1.3                           & 0.84                          \\
    Spin-down Luminosity ($10^{32}\; \erg\, \ps$) \dotfill & 0.30                          & 0.53                          & 0.67                          \\
    Characteristic Age (Myr)                      \dotfill & 8.5                           & 11                            & 15                            \\
    $820\; \MHz$ FWHM                             \dotfill & 0.019                         & 0.015                         & 0.015                         \\
    $820\; \MHz$ Flux Density ($\mJy$)            \dotfill & 0.4                           & 0.5                           & 0.4                           \\
    Spectral Index                                \dotfill & $-1.8$                        & $-2.3$                        & \nodata                       \\
    Rotation Measure ($\mathrm{rad}\, \pmsq$)     \dotfill & \nodata                       & 34.7(3.2)                     & $-47.6(6.0)$                  \\
  \enddata
\end{deluxetable}
\setcounter{table}{1}
\begin{deluxetable}{lcc}
  \centering
  \tabletypesize{\footnotesize}
  \tablewidth{0pt}
  \tablecolumns{3}
  \tablecaption{Parameters of Newly Discovered Long Period Pulsars (Cont.)}
  \tablehead{
    \colhead{Parameter} &
    \colhead{PSR J2013$-$0649} &
    \colhead{PSR J2033+0042}}
  \startdata
    \cutinhead{Timing Parameters}
    Right Ascension (J2000)                       \dotfill & 20:13:17.7507(38)             & 20:33:31.11(12)               \\
    Declination (J2000)                           \dotfill & $-$06:49:05.39(32)            & 00:42:22.0(8.0)               \\
    Pulsar Period ($\s$)                          \dotfill & 0.5801872690010(34)           & 5.01339800063(90)             \\
    Period Derivative ($\s\, \ps$)                \dotfill & 6.007(24)$\times$10$^{-16}$   & 1.013(78)$\times$10$^{-14}$   \\
    Dispersion Measure (\dmu)                     \dotfill & 63.36(10)                     & 37.84(13)                     \\
    Reference Epoch (MJD)                         \dotfill & 55172.0                       & 55172.0                       \\
    Span of Timing Data (MJD)                     \dotfill & 55005--55339                  & 55005--55339                  \\
    Number of TOAs                                \dotfill & 45                            & 22                            \\
    RMS Residual ($\us$)                          \dotfill & 150                           & 2195                          \\
    Error Factor                                  \dotfill & 1.370                         & 4.680                         \\
    \cutinhead{Derived Parameters}
    Galactic Longitude (deg)                      \dotfill & 36.17                         & 45.88                         \\
    Galactic Latitude (deg)                       \dotfill & $-$21.29                      & $-$22.2                       \\
    DM Derived Distance (kpc)                     \dotfill & 3.0                           & 1.9                           \\
    Surface Magnetic Field ($10^{12}$ Gauss)      \dotfill & 0.60                          & 7.2                           \\
    Spin-down Luminosity ($10^{32}\; \erg\, \ps$) \dotfill & 1.2                           & 0.032                         \\
    Characteristic Age (Myr)                      \dotfill & 15                            & 7.8                           \\
    $820\; \MHz$ FWHM                             \dotfill & 0.017                         & 0.018                         \\
    $820\; \MHz$ Flux Density ($\mJy$)            \dotfill & 0.6                           & 1.2                           \\
    Spectral Index                                \dotfill & \nodata                       & $-1.7$                        \\
    Rotation Measure ($\mathrm{rad}\, \pmsq$)     \dotfill & \nodata                       & $-71.2(2.2)$                  \\
  \enddata
  \tablecomments{Numbers in parentheses are 1-$\sigma$ uncertainties
  as determined by \texttt{TEMPO}; although we have scaled the TOA
  uncertainties by the error factors reported, we have not doubled the
  nominal \texttt{TEMPO} uncertainties as is sometimes done in these
  cases.  Flux density estimates typically have a 20--30\% relative
  uncertainty due to scintillation.  All timing solutions use the
  DE405 Solar System Ephemeris and the UTC(NIST) time system. Derived
  quantities assume an $R = 10\; \km$ neutron star with $I = 10^{45}\;
  \gm\, \cm^2$ \citep[see][]{lk05}.  The \dm\ derived distances were
  calculated using the NE2001 model of Galactic free electron density,
  and have typical errors of $\sim 20\%$ \citep{cl02}.}
\end{deluxetable}
\begin{deluxetable}{lcc}
  \centering
  \tabletypesize{\footnotesize}
  \tablewidth{0pt}
  \tablecolumns{3}
  \tablecaption{Parameters of Newly Discovered Short Period Pulsars \label{table:recycled_params}}
  \tablehead{
    \colhead{Parameter} &
    \colhead{PSR J0348+0432} &
    \colhead{PSR J1923+2515}}
  \startdata
    \cutinhead{Timing Parameters}
    Right Ascension (RA; J2000)                   \dotfill & 03:48:43.63817(33)            & 19:23:22.494560(76)           \\
    Declination (DEC; J2000)                      \dotfill & 04:32:11.449(10)              & 25:15:40.6436(14)             \\
    RA Proper Motion ($\mathrm{mas}\, \pyr$)      \dotfill & \nodata                       & $-$6.2(2.4)                   \\
    DEC Proper Motion ($\mathrm{mas}\, \pyr$)     \dotfill & \nodata                       & $-$23.5(7.0)                  \\
    Pulsar Period ($\s$)                          \dotfill & 0.039122656280156(10)         & 0.00378815551961303(52)       \\
    Period Derivative ($\s\, \ps$)                \dotfill & 2.417(16)$\times$10$^{-19}$   & 9.42(14)$\times$10$^{-21}$    \\
    Dispersion Measure (\dmu)                     \dotfill & 40.56(11)                     & 18.85766(19)                  \\
    Reference Epoch (MJD)                         \dotfill & 55278.0                       & 55322.0                       \\
    Span of Timing Data (MJD)                     \dotfill & 54873--55682                  & 55005--55639                  \\
    Number of TOAs                                \dotfill & 183                           & 153                           \\
    Error Factor                                  \dotfill & 1.657                         & 1.245                         \\
    RMS Residual ($\us$)                          \dotfill & 10.33                         & 5.0                           \\
    \cutinhead{Binary Parameters}
    Binary Model                                  \dotfill & ELL1                          & \nodata                       \\
    Orbital Period (days)                         \dotfill & 0.10242406134(30)             & \nodata                       \\
    Projected Semi-major Axis (lt-s)              \dotfill & 0.1409842(34)                 & \nodata                       \\
    Epoch of Ascending Node (MJD)                 \dotfill & 54889.70532337(65)            & \nodata                       \\
    1$^\mathrm{st}$ Laplace Parameter             \dotfill & $< 5.0 \times 10^{-5}$        & \nodata                       \\
    2$^\mathrm{nd}$ Laplace Parameter             \dotfill & $< 6.3 \times 10^{-5}$        & \nodata                       \\
    \cutinhead{Derived Parameters}
    Orbital Eccentricity                          \dotfill & $< 8.1 \times 10^{-5}$        & \nodata                       \\
    Mass Function (\Msun)                         \dotfill & 0.000286807(20)               & \nodata                       \\
    Minimum Companion Mass (\Msun)                \dotfill & 0.086                         & \nodata                       \\
    Galactic Longitude (deg)                      \dotfill & 183.34                        & 58.95                         \\
    Galactic Latitude (deg)                       \dotfill & $-$36.77                      & 4.75                          \\
    DM Derived Distance (kpc)                     \dotfill & 2.1                           & 1.6                           \\
    Transverse Velocity ($\km\, \ps$)             \dotfill & \nodata                       & 188(46)                       \\
    Shklovskii Effect ($\s\, \ps$)                \dotfill & \nodata                       & 8.9(4.7)$\times$10$^{-21}$    \\
    Intrinsic Spin-down ($\s\, \ps$)              \dotfill & \nodata                       & $< 5.3 \times 10^{-21}$\tablenotemark{a} \\
    Surface Magnetic Field ($10^9$ Gauss)         \dotfill & 3.1                           & $< 1.4$\tablenotemark{a}      \\
    Spin-down Luminosity ($10^{32}\; \erg\, \ps$) \dotfill & 1.6                           & $< 38$\tablenotemark{a}       \\
    Characteristic Age (Gyr)                      \dotfill & 2.6                           & $> 11$\tablenotemark{a}       \\
    $820\; \MHz$ FWHM                             \dotfill & 0.016                         & 0.142                         \\
    $820\; \MHz$ Flux Density ($\mJy$)            \dotfill & 1.8                           & 0.6                           \\
    Spectral Index                                \dotfill & $-1.2$                        & $-1.7$                        \\
    Rotation Measure ($\mathrm{rad}\, \pmsq$)     \dotfill & 49.5(13)                      & 10.8(3.8)                     \\
  \enddata
  \tablenotetext{a}{These quantities are limits based on our error for
  $\dot{P}\rmsub{int}$ after correcting for the Shklovskii effect. See
  \S\ref{sec:J1923} for further details.} 
  \tablecomments{Numbers in parentheses are 1-$\sigma$ uncertainties
  as determined by \texttt{TEMPO}; although we have scaled the TOA
  uncertainties by the error factors reported, we have not doubled the
  nominal \texttt{TEMPO} uncertainties as is sometimes done in these
  cases. Flux density estimates typically have a 20--30\% relative
  uncertainty due to scintillation.  All timing solutions use the
  DE405 Solar System Ephemeris and the UTC(NIST) time system. Derived
  quantities assume an $R = 10\; \km$ neutron star with $I = 10^{45}\;
  \gm\, \cm^2$ \citep[see][]{lk05}.  Minimum companion masses were
  calculated assuming a $1.4\; \Msun$ pulsar. The \dm\ derived
  distances were calculated using the NE2001 model of Galactic free
  electron density, and have typical errors of $\sim 20\%$
  \citep{cl02}.}
\end{deluxetable}

\end{document}